\documentclass[11pt]{article}
\usepackage{latexsym}
\usepackage{amsmath,amsfonts,amsbsy,amssymb,amscd}
\usepackage{fancybox,fancyhdr}
\usepackage{cite}
\usepackage{pstricks,pst-node,psfrag}
\usepackage{graphicx,epsfig,picture,graphics}

\psset{unit=1.3cm,linewidth=.5pt,radius=.2}  % controls the size of diagrams

\usepackage{float}                          % to force floaters to stay Here
\usepackage{lscape}                         % landscape laid out pages
\usepackage{bm}
\usepackage{tikz}
\newcommand{\hoch}[1]{$\, ^{#1}$}

\usepackage{color}
\definecolor{MyDarkBlue}{rgb}{0.15,0.15,0.45}    \definecolor{MyGreen}{rgb}{0.15,0.45,0.45}
\definecolor{MyPurple}{rgb}{0.55,0.25,0.55}
\usepackage[linktocpage=true]{hyperref}
\hypersetup{colorlinks=true,citecolor=MyGreen,linkcolor=MyPurple,urlcolor=MyGreen}

\begin{document}

\begin{flushright}
% \vspace*{1cm}
\hfill{ UG-15-68 \\ TUW-15-16}
\end{flushright}
\vskip 4cm

\begin{center}
{\Large \bf Newton--Cartan (super)gravity as a non-relativistic limit}
\end{center}
\vspace{25pt}
\begin{center}
{\Large {\bf }}

\vspace{15pt}

{\Large Eric Bergshoeff\hoch{1}, Jan Rosseel\hoch{2} and Thomas Zojer\hoch{1}}

\vspace{30pt}

\hoch{1} {\it Van Swinderen Institute for Particle Physics and Gravity, University of Groningen,\\
Nijenborgh 4, 9747 AG Groningen, The Netherlands}\\

\vspace{5pt}

\texttt{e.a.bergshoeff@rug.nl, t.zojer@rug.nl} \\

\vspace{10pt}

\hoch{2} {\it Institute for Theoretical Physics, Vienna University of Technology,\\
                               Wiedner Hauptstr.~8--10/136, A-1040 Vienna, Austria}\\

\vspace{5pt}

\texttt{rosseelj@hep.itp.tuwien.ac.at}

\vspace{45pt}
% \underline{ABSTRACT}
\end{center}

\noindent
We define a procedure that, starting from a relativistic theory of
supergravity, leads to a consistent, non-relativistic version
thereof. As a first application we use this limiting procedure to show
how the Newton--Cartan formulation of non-relativistic gravity can be
obtained from general relativity. Then we apply it in a supersymmetric
case and derive a novel, non-relativistic, off-shell formulation of
three-dimensional Newton--Cartan supergravity.

\vspace{15pt}

\thispagestyle{empty}

\vspace{15pt}

 \vfill

\voffset=-40pt

\newpage

\tableofcontents

% \addtocontents{toc}{\protect\setcounter{tocdepth}{2}}

% \newpage

% \thispagestyle{fancy}

%%%%%%%%%%%%%%%%%%%%%%%%%%%%%%%%%%%%%%%%%%%%%%%%%%%%%%%%%%%%%%%%%%%%%%%%%%%%%%

%%%%%%%%%%%%%%%%
%%%%%%%%%%%%%%%%                      Cut here
%%%%%%%%%%%%%%%%

\section{Introduction}

Recent advances in the study of non-relativistic field theories have
spurred a renewed interest in Newton--Cartan geometry. Originally
devised as a topic in differential geometry that treats Newtonian
space-times with a notion of absolute time and space, Newton--Cartan
geometry has mostly been discussed in the context of Newton--Cartan
gravity \cite{Cartan1,Cartan2}. The latter refers to a generally covariant and
geometric reformulation of Newtonian gravity that mimics General
Relativity as much as possible (see Chapter 12 of \cite{Misner:1974qy}
for a textbook exposition). Newton--Cartan geometry then plays a
similar role for Newton--Cartan gravity as Riemannian geometry does
for General Relativity.
\footnote{\label{foo1}
Non-relativistic gravitational theories are not unique, for example the gravitational
background that is used to describe the Newton--Cartan point-particle is different from the
background that describes non-relativistic branes, see e.g.~\cite{Andringa:2012uz}.
In this work we only consider backgrounds of the first kind, i.e.~'particle' backgrounds,
which feature more prominently in the literature.}

Recently, Newton--Cartan geometry has also been considered in entirely
different contexts. In particular, it has been crucial in work by Son
and collaborators on the fractional quantum Hall effect
\cite{Hoyos:2011ez,Son:2013rqa,Geracie:2014nka}. Here, Newton--Cartan
geometry and diffeomorphism invariance are used as a guiding principle
to construct an effective action for an external gauge-field and a
background metric source, that can be used to find the electromagnetic and
gravitational response of a quantum Hall fluid. It was argued that
this effective action can capture universal features of
the quantum Hall effect, other than the quantized Hall conductivity
that is determined by its lowest order Chern--Simons term. For
instance, the effective action naturally includes a so-called Wen--Zee
term \cite{Wen:1992ej}, that describes a coupling between the
gauge-field and spatial curvature and that universally encodes the
Hall viscosity.

Newton--Cartan geometry has also been instrumental in the context of
Lifshitz holography, that attempts to define a gravitational dual for
non-relativistic field theories that are invariant under time and
spatial translations, spatial rotations and anisotropic
dilatations. The putative gravitational dual is formulated around a
so-called `asymptotically locally Lifshitz' space-time and the
boundary geometry of such a space-time is described by Newton--Cartan
geometry with torsion \cite{Christensen:2013lma}. This observation
regarding the boundary geometry of Lifshitz space-times was
subsequently used in
\cite{Christensen:2013rfa,Hartong:2014oma,Hartong:2014pma,Hartong:2015wxa} to
e.g.~define the boundary stress tensor and to calculate holographic
Ward identities in Lifshitz holography.

The above developments have motivated recent studies on how
non-relativistic field theories can be appropriately coupled to
arbitrary Newton--Cartan backgrounds (with or without torsion)
\cite{Son:2005rv,Banerjee:2014nja, Jensen:2014aia,Jensen:2014ama,
  Geracie:2015xfa,Geracie:2015dea}. In most of the studies that
appeared in the literature, Newton--Cartan geometry is considered in a
metric formalism, that features two degenerate metrics of rank 1 and
rank $d$ (with $d$ the number of spatial dimensions), used to measure
temporal and spatial distances respectively. In such a formalism,
parallel transport is defined via an affine connection, that can be
defined via metric compatibility. While the metric formulation
provides a clear definition of Newton--Cartan geometry, for many
practical applications an equivalent vielbein formulation is often
more suitable. Such a vielbein formulation introduces local spatial
rotations and Galilean boosts as well as associated spin-connections
and can thus be very useful e.g.~when considering couplings to
fermions or incorporating supersymmetry.

At first
sight, studying supersymmetric non-relativistic field theories on arbitrary
Newton--Cartan backgrounds might seem rather academic. In the relativistic case
however, there exist powerful localization techniques that allow one to extract
exact results for supersymmetric
field theories on curved backgrounds \cite{Pestun:2007rz} (see also the
lecture notes \cite{Marino:2011nm} and references therein). A convenient
procedure for determining on which curved backgrounds supersymmetric theories
can be formulated and what supersymmetry algebra such theories exhibit, was
given in \cite{Festuccia:2011ws}. This procedure essentially consists of
coupling the flat space theory to off-shell supergravity, choosing a classical
background (specified by the metric and arbitrary values for the auxiliary
fields in the off-shell multiplet) and sending the Planck mass to infinity. If
one wishes to investigate whether these localization techniques can be extended
to non-relativistic field theories, one thus needs to obtain off-shell
realizations of non-relativistic supergravity. For this, an appropriate
vielbein formulation
of Newton--Cartan geometry is essential.

Torsionless Newton--Cartan geometry, as it appears in
Newton--Cartan gravity, was developed in terms of vielbeine in
\cite{Andringa:2010it,Andringa:2013mma}. It was in particular shown that this
geometry
can be obtained via a gauging of the Bargmann algebra, i.e.~the
central extension of the Galilei algebra. In this gauging, one introduces gauge-fields
for every generator of the Bargmann algebra, along with constraints on
some of the gauge covariant curvatures. The latter are interpreted as torsion
conditions
or as constraints that allow one to express local time and spatial
translations as diffeomorphisms. The temporal and spatial vielbeine then appear
as the gauge-fields associated to time and spatial translations and their
transformation rules are determined by the Bargmann algebra. Crucially, the
vielbein formulation obtained in this way also includes an independent
one-form that is interpreted as a gauge-field for central charge
transformations. The role of this extra central charge gauge-field in
constructing couplings to arbitrary curved non-relativistic backgrounds has been
discussed recently in \cite{Geracie:2015xfa,Geracie:2015dea}. The gauging
procedure has been extended to obtain an on-shell supergravity version of
three-dimensional Newton--Cartan gravity in \cite{Andringa:2013mma}. In
\cite{Bergshoeff:2014uea}, it was moreover shown that the torsional
Newton--Cartan geometry that appears in Lifshitz holography can be obtained from
gauging the Schr\"odinger algebra, a conformal extension of the Bargmann
algebra. Also in that case, the inclusion of a central charge gauge-field is
necessary.

While the gauging procedure provides an effective tool to construct a
vielbein formulation of bosonic Newton--Cartan geometry and gravity,
it is not always sufficient to obtain supersymmetric generalizations
thereof. In those cases, the vielbeine and central charge gauge-fields
are part of a supermultiplet, that might contain extra fields, that
cannot be interpreted as gauge-fields of an underlying
non-relativistic superalgebra. This is in particular, but not
exclusively, the case when off-shell formulations are considered. When
considering supersymmetry, a different procedure to obtain the field
content and transformation rules of a supermultiplet that contains the
vielbeine and the central charge gauge-field is therefore
necessary. The aim of this paper is to provide such a procedure and to
illustrate it in various examples.

The procedure developed here essentially consists of taking a
non-relativistic limit of vielbein formulations of relativistic (super)gravity.
We will show how such a limit can be defined and implemented in a consistent
manner. This limit in particular sheds light on how Newton--Cartan gravity in
the vielbein formulation can be obtained as a non-relativistic limit of General
Relativity. Apart from elucidating how Newton--Cartan geometry descends from
relativistic Riemannian geometry (see
\cite{Dautcourt,Kuenzle,Kuchar:1980tw,Jensen:2014wha,Geracie:2015dea} for early and more
recent work on this), the limiting procedure also has the advantage that it can
be used to obtain versions of Newton--Cartan geometry and gravity that cannot be
obtained via gauging. We will in particular use it to obtain an
off-shell version of the three-dimensional Newton--Cartan supergravity theory
constructed in \cite{Andringa:2013mma}. We should stress that the limiting
procedure we discuss in this work can be used to obtain versions of Newton--Cartan geometry, that
descend from relativistic geometries but that it is a priori not clear that
every non-relativistic geometry can be obtained in this way. The torsional
Newton--Cartan geometry constructed in \cite{Bergshoeff:2014uea} for instance is
based on the Schr\"odinger algebra, that cannot be obtained via a contraction
of a relativistic conformal algebra. This geometry might thus furnish an
example of a non-relativistic geometry that does not descend from a
relativistic one in an easy manner.

\bigskip\noindent This paper is organized as follows. In section
\ref{sec:procedure} we explain the general procedure that allows us to
obtain non-relativistic geometries from relativistic ones. We
illustrate this method with several examples that have been
constructed in the literature using other methods. As a first example,
we will show how the vielbein formulation of torsionless
Newton--Cartan geometry of \cite{Andringa:2010it} can be obtained in
this way. Our second example will deal with the on-shell
three-dimensional Newton--Cartan supergravity theory of
\cite{Andringa:2013mma}. As a third example, we will show how
non-relativistic particle actions can be obtained from relativistic
ones, using the limiting procedure. In section
\ref{sec:offshellsugra}, we will use the same procedure to obtain a
supersymmetric generalization of Newton--Cartan gravity that has not
yet appeared in the literature, namely an off-shell formulation of
three-dimensional Newton--Cartan supergravity. Finally, we conclude in
section \ref{sec:conclusions} and give an outlook on various problems
that could be handled using the methods described in this paper.

%%%%%%%%%%%%%%%%%%%%%%%%%%%%%%%%%%%%%%%%%%%%%%%%%%%%%%%%%%%%%%%%%%%%%%%%%%%%%%%%%%%%%%%%%%%%%%%%%%%%%%%%%%%%%
%%%%%%%%%%%%%%%%%%%%%%%%%%%%%%%%%%%%%%%%%%%%%%%%%%%%%%%%%%%%%%%%%%%%%%%%%%%%%%%%%%%%%%%%%%%%%%%%%%%%%%%%%%%%%

\section{The road to non-relativistic supergravity}\label{sec:procedure}

In this section we discuss how to derive non-relativistic geometries and
gravity theories from relativistic ones. We first describe our
procedure, which amounts to taking a non-relativistic limit in a consistent
manner, in section \ref{subsec:genproc}. We then illustrate this limiting
procedure in three examples.
In subsection \ref{subsec:NC} we re-derive the results of \cite{Andringa:2010it}
regarding the vielbein description of torsionless Newton--Cartan gravity in
arbitrary dimensions and in subsection \ref{subsec:onshellNC}
we re-derive the on-shell three-dimensional Newton--Cartan supergravity theory
of \cite{Andringa:2013mma}. In subsection \ref{subsec:curvpp}, we apply the
limiting procedure to derive the results obtained in
\cite{Bergshoeff:2014gja} for the
non-relativistic superparticle in a curved background.

%%%%%%%%%%%%%%%%%%%%%%%%%%%%%%%%%%%%%%%%%%%%%%%%%%%%%%%%%%%%%%%%%%%%%%%%%%%%%%%%%%%%%%%%%%%%%%%%%%%%%%%%%%%%%%

\subsection{The general procedure}\label{subsec:genproc}

The method used in this paper can be viewed as an
extension of the contraction of a relativistic space-time symmetry
algebra to a non-relativistic one. In particular, its aim is to mimic
the algebra contraction to obtain an irreducible multiplet of fields
that represents the contracted non-relativistic algebra starting from an
irreducible multiplet of the parent relativistic algebra.

Recall that when performing a standard In\"on\"u--Wigner contraction
of a symmetry algebra one first redefines the generators of the
algebra, by taking linear combinations of the original generators with
coefficients that involve a contraction parameter $\omega$. The
contracted algebra is then obtained by calculating commutators of the
redefined generators, re-expressing the result in terms of them and
taking $\omega \rightarrow \infty$ in the end. An In\"on\"u--Wigner
contraction performed in this way does not change the number of
generators. Moreover, when considering finite $\omega$ the algebra of
redefined generators is equivalent to the original one.

When extending this algebra contraction to an irreducible multiplet of
fields that represents the parent relativistic algebra, it is useful
to divide the fields in three categories. A first category consists of
independent fields that can be viewed as gauge-fields that are
associated to certain generators of the algebra. For instance, the
vielbein of General Relativity roughly plays the role of the
gauge-field of local translations \cite{Chamseddine:1976bf}. A second
category comprises gauge-fields that are not independent, but that
instead depend on other fields in the multiplet. This
is the way in which the spin-connection of General Relativity can be
viewed, namely  as a dependent gauge-field for local Lorentz transformations. Finally, the
last category contains independent fields that cannot be interpreted
as gauge-fields of the underlying space-time symmetry algebra. This is
for instance the case when considering off-shell supergravity
multiplets, where typically auxiliary fields are necessary to guarantee
the closure of the commutator algebra and, in the relativistic case, to ensure
that the number of bosonic and fermionic degrees of freedom match.
For simplicity, we will call all such additional fields, that are not gauge-fields corresponding to a generator,
auxiliary fields.
The fields that can be viewed as gauge-fields corresponding
to the generators of the algebra  are in general subject to
constraints. Some of these constraints are called `conventional' and
merely serve to express the dependent gauge-fields in terms of the
independent ones. These constraints are identically satisfied, once the explicit
expressions for the dependent fields are plugged in. There will in
general also be a second type of `un-conventional' constraints, that are
not identically satisfied. This second type of constraints will play a crucial role in
ensuring consistency of the limiting procedure.

The first step of the limiting procedure consists of extending the
redefinition of the algebra generators, with the contraction parameter
$\omega$, to all fields and symmetry parameters of the relativistic multiplet. Let us
therefore denote the parent algebra generators collectively by $T_A$,
the parent symmetry parameters by $\xi^A$ and the parent fields that
correspond to gauge-fields by $A_\mu^A$. The redefinition of the
generators $T_A$ to generators $\tilde{T}_A$, that involves $\omega$
and defines the contraction, can then be extended to redefinitions of
$\xi^A$ to $\tilde{\xi}^A$ and of $A_\mu^A$ to $\tilde{A}_\mu^A$ such
that\,\footnote{Strictly speaking, the second equation only holds up to terms that are subleading in $\omega$.
This will be clarified in the specific examples, see e.g.~the discussion leading to eq.~\eqref{eq26}.}
\begin{equation} \label{redefparfields}
 \xi^A T_A = \tilde{\xi}^A \tilde{T}_A \,, \qquad A_\mu^A T_A = \tilde{A}_\mu^A
\tilde{T_A} \,.
\end{equation}
This defines the tilded parameters and fields in terms of the original
ones and the contraction parameter $\omega$. For finite $\omega$, this
redefinition merely corresponds to a field redefinition and the
redefined multiplet is equivalent to the original one. Note that
\eqref{redefparfields} involves both independent and dependent
gauge-fields. For the dependent fields, one should take special care
that the redefinition suggested by \eqref{redefparfields} is
consistent with the one obtained by performing the redefinitions in
the explicit expressions of the dependent fields in terms of the
independent ones and that their $\omega \rightarrow \infty$ limit is
well-defined. This amounts to a non-trivial consistency check, that we
will discuss further in the next paragraph. Equation
\eqref{redefparfields} does not determine the redefinitions of the
auxiliary fields. These are found by examining all transformation
rules in terms of redefined gauge-parameters and gauge-fields and by
requiring that no term in the transformation rules diverges when
taking the limit $\omega \rightarrow \infty$. As we will see in
specific examples, this can typically be achieved by suitably
rescaling the auxiliary fields with the contraction parameter
$\omega$.

Once all redefinitions have been determined, one can apply them in the
transformation rules of all independent fields and in the
un-conventional constraints. Their non-relativistic versions are then
obtained by sending $\omega \rightarrow \infty$. For properly chosen
redefinitions, no divergences are encountered here. One does however
need to check whether other quantities are finite in this limit. In
particular, one needs to examine the expressions of the dependent
fields in terms of the independent ones and check whether one obtains
a finite result, consistent with the redefinition implied by
\eqref{redefparfields}, in the limit $\omega \rightarrow
\infty$. Typically, terms that are proportional to positive powers of
$\omega$, and hence blow up in the $\omega\to\infty$ limit,  do show up in the expressions
for the dependent gauge-fields. One can however use the
un-conventional constraints, written in terms of redefined fields, to
replace these by terms that are finite or vanishing in the $\omega
\rightarrow \infty$ limit. In this way, the relativistic dependent
gauge-fields have a well-defined $\omega \rightarrow \infty$ limit,
consistent with \eqref{redefparfields} and lead to the correct finite expressions for the
non-relativistic dependent gauge fields.

Once all non-relativistic transformation rules, dependent gauge-fields and
constraints have been found in this way, one needs to check whether
the constraints form a consistent set. This involves varying all
non-trivial constraints found so far under all symmetry
transformations and checking that they form a closed set.

Finally, we mention that the limiting procedure can lead to the
elimination of a number of auxiliary fields. This is due to the fact that we are
interested in obtaining an irreducible multiplet. Loosely speaking, the
non-relativistic theory can have less equations of motion than the
relativistic one. The number of auxiliary fields that are needed to
realize the non-relativistic algebra can therefore differ from the
number that is needed to realize the relativistic algebra. This explains
why  some of the auxiliary fields can be eliminated in the
limiting process.

We can summarize the procedure in the following way:
\begin{itemize}
\item[I.] We first write the relativistic gauge-fields in terms of new
  redefined ones, using a contraction parameter $\omega$. This field redefinition
  is dictated by the same redefinition that one performs on  the generators to define the contraction of the algebra.
  The new fields will become the
  proper non-relativistic fields after we have taken the $\omega\to\infty$ limit. At this point the
  scaling of the auxiliary fields is still arbitrary.
\item[II.] Using the above redefinitions and taking the limit $\omega \rightarrow \infty$
we can derive a first set of
  non-relativistic constraints by taking the limit $\omega\rightarrow \infty$ of the relativistic un-conventional ones.
\item[III.] In a third step we derive the transformation rules of all
  fields. Requiring that no terms diverge in the limit $\omega
  \rightarrow \infty$ fixes the scalings of the auxiliary fields. At
  this point, we can check the limit of the dependent gauge-fields, such
  as, e.g., the spin-connection field. Requiring that they have a well-defined
  limit may involve the use of the un-conventional constraints, written in
  terms of the redefined fields, in order to replace divergent terms by terms with
  a proper limit.
\item[IV.] In this step we check whether the constraints found in step II
form a closed set under the different symmetry transformations or whether we are forced to
introduce additional constraints.
An example where many new constraints are found by continuous variation under supersymmetry is given by
the chain of constraints in eqs.~\eqref{moreconstraints}--\eqref{onshell}.
\item[V.] The number of auxiliary fields that are needed in the non-relativistic case may be less
then the number that is needed in the relativistic case. In such cases, in order to obtain an irreducible multiplet,
we eliminate the redundant auxiliary fields. This occurs, for instance, in the example of section \ref{sec:offshellsugra}.
\end{itemize}

In the next subsections, we will illustrate the above limiting  procedure by applying it
to re-derive various results on Newton--Cartan geometry and
(super)gravity that have been obtained in the literature using  other methods.

%%%%%%%%%%%%%%%%%%%%%%%%%%%%%%%%%%%%%%%%%%%%%%%%%%%%%%%%%%%%%%%%%%%%%%%%%%%%%%%%%%%%%%%%%%%%%%%%%%%%%%%%%%%%%%

\subsection{Example 1: Newton--Cartan geometry and gravity
from General Relativity}\label{subsec:NC}

In this section we illustrate the limiting procedure that we just
described, to obtain the vielbein description of Newton--Cartan
gravity of \cite{Andringa:2010it} from the vielbein formulation of
General Relativity. We will pay specific attention to how the
transformation rules of the Newton--Cartan  fields arise, to
how the limiting procedure leads to the correct constraints that these
 fields have to obey and to how the correct dependent
gauge-fields are obtained. First, we summarize in section
\ref{subsubsec:gr} our starting point, namely the kinematics of
General Relativity as a gauging of the Poincar\'e algebra. Then, using
those formulas, we deduce Newton--Cartan geometry and gravity in
section \ref{subsubsec:NC}.

\subsubsection{The kinematics of General Relativity}\label{subsubsec:gr}

It is well-known that the vielbein formulation of General Relativity
can be viewed as a gauging of the Poincar\'e algebra
\cite{Chamseddine:1976bf}. Here we will briefly summarize this gauging
procedure. Note that this leads to the kinematics of General
Relativity. In order to obtain the dynamics, one has to supplement the
formulas collected here with the Einstein equations. If one is however
only interested in geometrical aspects and not so much in dynamics,
one need not do so. This section will serve as the starting point of
our limiting procedure, that will lead to the formulation of
Newton--Cartan gravity as obtained via a gauging of the Bargmann
algebra in \cite{Andringa:2010it}.

The kinematics of General Relativity in a $(d+1)$-dimensional spacetime is described by the vielbein $E_\mu{}^A$
and spin-connection $\Omega_\mu{}^{AB}$, $A,B = 0,\cdots, d$, that can be
viewed as gauge-fields associated to translations $\hat P_A$ and Lorentz rotations
$M_{AB}$. Under diffeomorphisms (with parameter $\xi^\mu$) and Lorentz
rotations (with parameter $\lambda^{AB}$) these fields transform as
\begin{align}
 \delta E_\mu{}^A &= \xi^\rho\partial_\rho E_\mu{}^A +E_\rho{}^A\partial_\mu\xi^\rho
                     +\lambda^A{}_B E_\mu{}^B \,, \label{reletrafo}\\[.1truecm]
 \delta \Omega_\mu{}^{AB} &= \xi^\rho\partial_\rho \Omega_\mu{}^{AB} +\Omega_\rho{}^{AB}\partial_\mu\xi^\rho
                   + \partial_\mu \lambda^{AB} +2\,\lambda^{[A}{}_C\,\Omega_\mu{}^{CB]} \label{deltaOmega}\,.
\end{align}
One may define gauge covariant curvatures, dictated by the structure constants
of the Poincar\'e algebra, as follows:
\begin{align}
 R_{\mu\nu}{}^A(E) &= 2\,\partial_{[\mu}E_{\nu]}{}^A -2\,\Omega_{[\mu}{}^A{}_B\,E_{\nu]}{}^B \,, \label{torsion} \\[.1truecm]
 R_{\mu\nu}{}^{AB}(\Omega) &= 2\,\partial_{[\mu}\Omega_{\nu]}{}^{AB} -2\,\Omega_{[\mu}{}^A{}_C\,\Omega_{\nu]}{}^{CB}
            \label{relcurv}\,.
\end{align}
The spin-connection $\Omega_\mu{}^{AB}$ is not an independent field; rather it
is given in terms of $E_\mu{}^A$ by solving the
torsion constraint
\begin{align} \label{reltorsionconstraint}
R_{\mu\nu}{}^A(E)=0\,.
\end{align}
This constraint also allows one to replace local translations
by general coordinate transformations.
The solution of \eqref{reltorsionconstraint} is given by
\begin{align}\label{relomega}
 \Omega_\mu{}^{AB}(E)=-2\,E^{\rho[A}\partial_{[\mu} E_{\rho]}{}^{B]}
                   +E_{\mu C} E^{\rho A} E^{\nu B} \partial_{[\rho} E_{\nu]}{}^C \,.
\end{align}
Note that imposing the torsion constraint \eqref{reltorsionconstraint} also
implies that the Riemann curvature tensor \eqref{relcurv} identically satisfies
the Bianchi identity
\begin{align}\label{relBianchi}
 R_{[\mu\nu\rho]}{}^B\big(\Omega(E)\big)=R_{[\mu\nu}{}^{AB}\big(\Omega(E)\big)\,E_{\rho]A}=0 \,.
\end{align}
The above defines the kinematics of General Relativity. The dynamics can be obtained
by imposing equations of motion, i.e.~putting the theory on-shell. In General
Relativity this is done by imposing the Einstein equations
\begin{align}\label{Einsteineq}
 E^\mu{}_A \,R_{\mu\nu}{}^{AB}\big(\Omega(E)\big)=0 \,.
\end{align}

The limiting procedure by which we will
obtain non-relativistic geometry and gravity is an extension of contractions of
relativistic symmetry algebras. As explained in \cite{Andringa:2010it},
torsionless Newton--Cartan geometry and gravity is linked to the Bargmann
algebra. In order to obtain the Bargmann algebra from an In\"on\"u--Wigner
contraction, one should start from a direct sum of the Poincar\'e algebra
with an abelian factor with generator $\cal{Z}$. In this way, one ensures that
the algebra before contraction has the same number of generators as the
Bargmann algebra. The abelian factor is represented by  an abelian gauge-field $M_\mu$,
that transforms under diffeomorphisms and abelian gauge transformations (with
parameter $\Lambda$) as follows:
\begin{align}\label{Mtransf}
 \delta M_\mu = \xi^\rho \partial_\rho M_\mu + M_\rho\, \partial_\mu \xi^\rho +
\partial_\mu \Lambda \,.
\end{align}
The curvature of $M_\mu$ is given by
\begin{align}\label{Mcurv}
 F_{\mu\nu}(M) = 2\,\partial_{[\mu}M_{\nu]} \,.
\end{align}
In order to take the non-relativistic limit we will need to impose
constraints on this curvature. For example, if we consider the
dynamics of General Relativity, we do not want to add extra degrees of
freedom, apart from the ones contained in the vielbein. In this case,
we will thus set the curvature $F_{\mu\nu}(M)$ to zero so that $M_\mu$
corresponds to a pure gauge-field. Another example where we will
constrain $F_{\mu\nu}(M)$ to be zero, will appear when we discuss
on-shell supersymmetry. There, this constraint will ensure that the
equality of bosonic and fermionic degrees of freedom in the
relativistic multiplet is not upset, after $M_\mu$ is added by hand to
an existing on-shell multiplet.
Even if we are only interested in the kinematics we cannot allow for a
completely arbitrary $F_{\mu\nu}(M)$. As we will see in the following
we have to constrain the spatial projection of $F_{\mu\nu}(M)$ to be zero
in order to take the non-relativistic limit in a consistent manner. In
particular, this will be crucial to obtain finite expressions for the
non-relativistic dependent spin-connections as limits of the
relativistic one.

In the following section, we will apply the limiting procedure to the formulas
collected above.

\subsubsection{Newton--Cartan gravity from relativistic
gravity}\label{subsubsec:NC}

In this first example, we will extend the contraction that gives the Bargmann
algebra from the Poincar\'e algebra, to the
vielbein and spin-connection of General Relativity, along the lines of section
\ref{subsec:genproc}. As explained above, the contraction and its extension
involve redefining generators and fields using a contraction
parameter $\omega$. We will be careful in distinguishing quantities that are
merely redefined relativistic ones, for which $\omega$ is finite, from
non-relativistic ones, that are obtained in the limit $\omega \rightarrow
\infty$. In particular, we will denote the former ones with a tilde, whereas for
the latter the tilde will be dropped.

Let us first briefly recall the
In\"on\"u--Wigner contraction of the Poincar\'e algebra to the Bargmann
algebra. The contraction is best
described by starting from a direct sum of the Poincar\'e algebra
(with translation generators $\hat{P}_A$ and Lorentz generators $M_{AB}$)
with an abelian factor (with generator $\cal{Z}$).\footnote{We have indicated the relativistic
translation generators with a hat to distinguish them from the redefined generators (indicated
with a tilde) and the Bargmann generators (with no hat or tilde).} Starting from the Poincar\'e
algebra
\begin{align}\label{Poincarealg}
 \big[\hat P_A,M_{BC}\big]=2\,\eta_{A[B}\,\hat P_{C]} \,, \hskip2cm
 \big[M_{AB},M_{CD}\big]=4\,\eta_{[A[C}\,M_{D]B]} \,,
\end{align}
supplemented with the generator $\mathcal{Z}$, we redefine the
generators using a contraction parameter $\omega$ as follows
\footnote{\label{foo4}Note that this redefinition corresponds to a 
non-relativistic particle limit where time is singled out as a special direction. One can  define more general $p$-brane limits where 
one time and $p$ spatial directions are singled out, see e.g.~\cite{Gomis:2000bd,Danielsson:2000gi,Gomis:2004pw,Andringa:2012uz}
and the comment in footnote \ref{foo1}.}
\begin{align}\label{contraction}
 \hat P_0\to \frac{1}{2\omega}\,\tilde H +\omega\,\tilde Z \,,
 \hskip1.5cm \mathcal{Z}\to \frac{1}{2\omega}\,\tilde H -\omega\,\tilde Z \,,
 \hskip1.5cm M_{a0}\to\omega\,\tilde G_a \,,
\end{align}
where we have split the space-time index $A$ into a time-like $0$-index and
spatial $a,b$-indices. Note that the spatial translations $\hat{P}_a$ and spatial
rotations $M_{ab}$ are left untouched, i.e.~$\hat{P}_a=\tilde P_a$ and $M_{ab}=\tilde M_{ab}$ .
We will in the following denote
$\tilde M_{ab} = \tilde J_{ab}$ to conform to earlier literature. Calculating
commutators of $\tilde H$, $\tilde P_a$, $\tilde G_a$, $\tilde J_{ab}$, $\tilde Z$, re-expressing the result
in terms of the same generators and taking
$\omega\to\infty$, we obtain the Bargmann algebra with the non-vanishing
commutators
\begin{align}\begin{aligned}\label{Bargmannalg}
 \big[P_a,J_{bc}\big] &= 2\,\delta_{a[b}\,P_{c]} \,,\hskip.6cm& \big[J_{ab},J_{cd}\big]&=4\,\delta_{[a[c}\,J_{d]b]} \,, &&\\[.1truecm]
 \big[G_a,J_{bc}\big] &= 2\,\delta_{a[b}\,G_{c]} \,, & \big[H,G_a\big] &= P_a \,, &
 \big[P_a,G_b\big] &= \delta_{ab}\,Z \,.
\end{aligned}\end{align}
In order to derive Newton--Cartan gravity, as obtained via the gauging of
\eqref{Bargmannalg} in \cite{Andringa:2010it}, we extend the contraction
\eqref{contraction} to the vielbein $E_\mu{}^A$, spin-connection
$\Omega_\mu{}^{AB}(E)$ and abelian gauge-field $M_\mu$ of the previous section.
In particular, we make the following redefinition for the relativistic
vielbein:
\begin{align}\label{Eexpand}
  E_\mu{}^A = \delta_0^A\,\big(\omega\,\tilde \tau_\mu+\frac{1}{2\omega}\, \tilde m_\mu\big) +\delta_a^A\,\tilde e_\mu{}^a \,,
\end{align}
where $\tilde\tau_\mu$, $\tilde e_\mu{}^a$ and $\tilde m_\mu$ will
be identified, in the limit $\omega\rightarrow \infty$ (where we will drop the tilde), as the independent gauge-fields of
Newton--Cartan geometry and
gravity. It is convenient to  define fields $\tilde e^\mu{}_a$, $\tilde \tau^\mu$  as follows:
\begin{align}\label{expressions}
 \tilde e^\mu{}_a\tilde e_\mu{}^b &= \delta_a^b \,, \hskip1cm  \tilde\tau^\mu\tilde\tau_\mu = 1 \,, \hskip1cm
 \tilde\tau^\mu \tilde e_\mu{}^a=\tilde\tau_\mu \tilde e^\mu{}_a = 0 \,, \hskip1cm  \tilde e^\rho{}_a\tilde e_\mu{}^a =
 \delta_\mu^\rho -\tilde\tau_\mu\tilde\tau^\rho \,.
\end{align}
In the limit $\omega\to\infty$, the fields $\tau_\mu\,,\tau^\mu\,, e_\mu{}^a$ and $e^\mu{}_a$ can be used
to define two separate  non-degenerate  Galilei-invariant metrics, one in the time direction and one in the three
spatial directions. These will not be needed in the present discussion.
Using the defining relations \eqref{expressions} the following expansion of the relativistic inverse vielbein is obtained
\begin{align}\label{Einvexpand}
 E^\mu{}_A = \delta^a_A\,\big[\tilde e^\mu{}_a +\mathcal{O}\big(\frac{1}{\omega^2}\big)\big]
            +\frac1\omega\,\delta^0_A\,\big[\tilde\tau^\mu +\mathcal{O}\big(\frac{1}{\omega^2}\big)\big] \,.
\end{align}
Note that in this expression, we have only explicitly given the terms of
leading order in $\omega$. There are in principle an infinite
number of corrections of lower order in $\omega$, that we have denoted by
$\mathcal{O}\big(1/\omega^2\big)$ and that will not be needed in the
following (as they will not contribute in the $\omega \rightarrow \infty$
limit).

The abelian gauge-field $M_\mu$ is redefined as follows
\begin{align}\label{Mlimit}
 M_\mu = \omega\,\tilde \tau_\mu - \frac{1}{2\omega}\,\tilde m_\mu \,.
\end{align}
The relativistic spin-connection $\Omega_\mu{}^{AB}(E)$ is a dependent field,
determined by the torsion constraint \eqref{reltorsionconstraint}.
As already mentioned under eq.~\eqref{Mcurv}, we now have to impose a constraint on
the curvature of the relativistic gauge-field $M_\mu$ in order to lower the powers of $\omega$ in certain terms
in the expression for $\Omega_\mu{}^{AB}(E)$, see eq.~\eqref{Mcurvexpl} below, such that  the limit can be taken.
In particular we mentioned the following choices
\begin{align}
 {\rm dynamical:}& \quad F_{\mu\nu}(M)=0 \,, \label{onMcurv} \\
 {\rm kinematical:}& \quad \tilde e^\mu{}_a \tilde e^\nu{}_b \,F_{\mu\nu}(M)=\tilde F_{ab}(M)=0  \label{offMcurv}\,.
\end{align}
The last choice is the least restrictive one and is sufficient to take
the $\omega \rightarrow \infty$ limit in a consistent manner, if one
is only interested in geometrical and kinematical aspects. The first
choice represents a stronger condition and should be adopted when one
is also interested in taking the non-relativistic limit of dynamical
aspects of General Relativity. In particular, this constraint implies
that $M_\mu$ is pure gauge and does not represent extra degrees of
freedom in the parent relativistic theory.

We can then define spin- and boost-connections
$\tilde\omega_\mu{}^{ab}(\tilde e,\tilde\tau,\tilde m)$,
$\tilde\omega_\mu{}^a(\tilde e,\tilde\tau,\tilde m)$, that will be
identified with the non-relativistic ones when $\omega \rightarrow
\infty$, as the coefficients of the terms of leading order in an
$\omega$-expansion of $\Omega_\mu{}^{AB}(E)$:
\begin{align}
 \Omega_\mu{}^{ab}(E) &= \tilde\omega_\mu{}^{ab}(\tilde e,\tilde\tau,\tilde m)
                   +\mathcal{O}\big(\frac{1}{\omega^2}\big) \,, \label{omegaabdef}\\[.1truecm]
 \Omega_\mu{}^{0a}(E) &= \frac1\omega\,\tilde\omega_\mu{}^a(\tilde e,\tilde\tau,\tilde m)
                   +\mathcal{O}\big(\frac{1}{\omega^3}\big) \,,
\label{omegaadef}
\end{align}
where
\begin{align}
\label{omegaab}
 \tilde\omega_\mu{}^{ab}(\tilde e,\tilde\tau,\tilde m) &= -2\,\tilde e^{\nu[a}\partial_{[\mu}\tilde e_{\nu]}{}^{b]}
      +\tilde e_\mu{}^c\tilde e^{\rho a}\tilde e^{\nu b}\partial_{[\rho}\tilde e_{\nu]}{}^c
      -\tilde\tau_\mu \tilde e^{\rho a}\tilde e^{\nu b}\partial_{[\rho}\tilde m_{\nu]} \,, \\[.1truecm]
\label{omegaa}
 \tilde\omega_\mu{}^a(\tilde e,\tilde\tau,\tilde m) &= \tilde\tau^\nu \partial_{[\mu}\tilde e_{\nu]}{}^a
      +\tilde e_\mu{}^b \tilde e^{\rho a}\tilde \tau^\nu \partial_{[\rho}\tilde e_{\nu]}{}^b
      +\tilde e^{\nu a}\partial_{[\mu}\tilde m_{\nu]}
      -\tilde\tau_\mu \tilde e^{\rho a}\tilde\tau^\nu \partial_{[\rho}\tilde m_{\nu]} \,.
\end{align}
Note that to obtain these formulas, one only needs the kinematical constraint \eqref{offMcurv} in the form
\begin{equation} \label{Mcurvexpl}
\omega \,\tilde e^\mu{}_a \tilde e^\nu{}_b \,\partial_{[\mu} \tilde\tau_{\nu]} =
        \frac{1}{2\omega}\,\tilde e^\mu{}_a \tilde e^\nu{}_b \, \partial_{[\mu} \tilde m_{\nu]} \,,
\end{equation}
to replace terms that diverge in the $\omega \to \infty$ limit by
terms that have the correct leading $\omega$-order as indicated in the
expansions \eqref{omegaabdef} and \eqref{omegaadef}. Since the
stronger constraint (\ref{onMcurv}) implies (\ref{Mcurvexpl}), it will
achieve the same goal.  The subleading terms in \eqref{omegaabdef} and
\eqref{omegaadef} are due to the fact that the relativistic
spin-connection $\Omega_\mu{}^{AB}(E)$ depends on the inverse vielbein
$E^\mu{}_A$.

The rationale behind the redefinitions \eqref{Eexpand}--\eqref{omegaadef} is
that they leave the sum of the products of the gauge-fields with their
respective generators invariant, up to subleading terms in $\omega$, that stem
from the dependent spin-connection via \eqref{omegaabdef} and
\eqref{omegaadef}. One thus has:
\begin{align}\label{eq26}
\begin{split}
 \hat P_A\,E_\mu{}^A& +\mathcal{Z}\,M_\mu +M_{AB}\,\Omega_\mu{}^{AB}(E) = \\[.1truecm]
   &=      \tilde P_a\,\tilde e_\mu{}^a +\tilde H\,\tilde\tau_\mu +\tilde Z\,\tilde m_\mu
         +\tilde J_{ab}\,\tilde \omega_\mu{}^{ab}(\tilde e,\tilde\tau,\tilde m)
       -2\,\tilde G_a\,\tilde \omega_\mu{}^a(\tilde e,\tilde\tau,\tilde m) +\mathcal{O}\big(\frac{1}{\omega^2}\big)\,.
\end{split}\end{align}

We proceed by taking the limit $\omega\to\infty$ and derive the kinematics of Newton--Cartan gravity. For example, dropping the
tildes on all fields we see that the expressions \eqref{omegaab} and \eqref{omegaa} coincide with the expressions of
\cite{Andringa:2010it}, which were obtained by setting
\begin{align}\begin{split}\label{convconstr}
 R_{\mu\nu}{}^a(P) &=2\,\partial_{[\mu}e_{\nu]}{}^a -2\,\omega_{[\mu}{}^{ab}e_{\nu]}{}^b
                     -2\,\omega_{[\mu}{}^a\tau_{\nu]} =0 \,, \\[.1truecm]
 R_{\mu\nu}(Z) &= 2\,\partial_{[\mu}m_{\nu]} -2\,\omega_{[\mu}{}^ae_{\nu]}{}^a =0 \,.
\end{split}\end{align}
These constraints are thus satisfied identically. Further constraints can be derived from the
relativistic Bianchi identity \eqref{relBianchi} and the constraints on the curvature of the relativistic
gauge-field $M_\mu$, see eq.~\eqref{onMcurv} or \eqref{offMcurv}. Using the inverse
vielbein \eqref{Einvexpand} and the expansion
\begin{align}
 R_{\mu\nu}{}^{AB}(\Omega)=\delta^A_a\delta^B_b\,\tilde R_{\mu\nu}{}^{ab}(\tilde J)
                           -\frac1\omega\,\delta^A_a\delta^B_0\,\tilde R_{\mu\nu}{}^a(\tilde G)
                           +\frac1\omega\,\delta^A_0\delta^B_b\,\tilde R_{\mu\nu}{}^b(\tilde G) \,,
\end{align}
in \eqref{relBianchi} we obtain the non-relativistic Bianchi identities
\begin{align}\label{nrBianchi}
 R_{[\mu\nu}{}^a(G)\,e_{\rho]}{}^a=0\,,\hskip1.5cm
      R_{[\mu\nu}{}^{ab}(J)\,e_{\rho]}{}^b +R_{[\mu\nu}{}^a(G)\,\tau_{\rho]} =0\,,
\end{align}
where the curvatures of the spin- and boost-connection gauge-fields are given by
\begin{align}\begin{split}
 R_{\mu\nu}{}^{ab}(J) &= 2\,\partial_{[\mu}\omega_{\nu]}{}^{ab}
              -2\,\omega_{[\mu}{}^a{}_c\,\omega_{\nu]}{}^{cb} \,, \\[.1truecm]
 R_{\mu\nu}{}^a(G) &= 2\,\partial_{[\mu}\omega_{\nu]}{}^a -2\,\omega_{[\mu}{}^{ab}\,\omega_{\nu]}{}^b \,.
\end{split}\end{align}
We consider the implications of eqs.~\eqref{onMcurv} and \eqref{offMcurv}
separately. In the first case (\ref{onMcurv}), we simply find
\begin{align}\label{Rhconstr}
 R_{\mu\nu}(H) = 2\,\partial_{[\mu}\tau_{\nu]} =0 \,.
\end{align}
In the kinematical case (\ref{offMcurv}), we find the less restrictive condition
\begin{align}\label{offRhconstr}
 R_{ab}(H) = 2\,e^\mu{}_ae^\nu{}_b\,\partial_{[\mu}\tau_{\nu]} =0 \,.
\end{align}
Interestingly, this constraint is equivalent to writing
\begin{align}\label{TTNC??}
 \partial_{[\mu}\tau_{\nu]} = b_{[\mu}\tau_{\nu]} \,,
\end{align}
where $b_\mu$ is completely arbitrary. This resembles the constraint found in the twistless torsional
Newton--Cartan geometry of \cite{Bergshoeff:2014uea}.

In either of the two cases, we can show that no further constraints are obtained by applying symmetry
transformations on the constraints \eqref{Rhconstr} and
\eqref{offRhconstr}. To do so
we need to derive the transformation rules of the Newton--Cartan
 fields $\tau_\mu$, $e_\mu{}^a$ and $m_\mu$. This can be
done by applying the relativistic transformation rules \eqref{reletrafo} and \eqref{Mtransf} to the
decompositions \eqref{Eexpand} and \eqref{Mlimit}.
For this purpose, we first express the new fields in
terms of the old ones, i.e.
\begin{align}\label{tauandm}
 \tilde\tau_\mu=\frac{1}{2\omega}\,\big(E_\mu{}^0 +M_\mu\big) \,, \hskip2cm \tilde m_\mu = \omega\,\big(E_\mu{}^0 -M_\mu\big) \,.
\end{align}
Having done this, it  is straightforward to obtain
\begin{align}\begin{split} \label{NCtraforules}
 \delta \tau_\mu &=0 \,, \\[.1truecm]
 \delta e_\mu{}^a &=\lambda^a{}_b \, e_\mu{}^b +\lambda^a\tau_\mu \,, \\[.1truecm]
 \delta m_\mu &=\partial_\mu\sigma +\lambda_a \,e_\mu{}^a \,,
\end{split}\end{align}
where we defined
\begin{align}
 \lambda^a=\omega\,\lambda^a{}_0 \,, \hskip2cm \Lambda=-\frac{\sigma}{\omega} \,,
\end{align}
in agreement with eqs.~\eqref{omegaadef} and \eqref{Mlimit}.
All fields transform under diffeomorphisms in the usual way.
The transformations of the spin-connect\-ions can be found as well:
\begin{align}\begin{split}\label{deltaomega2}
 \delta \omega_\mu{}^{ab}(e,\tau,m) &= \partial_\mu\lambda^{ab} +2\,\lambda^{[a}{}_c\,\omega_\mu{}^{cb]} \,, \\[.1truecm]
 \delta \omega_\mu{}^a(e,\tau,m) &= \partial_\mu \lambda^a +\lambda^a{}_b\,\omega_\mu{}^b
                         -\omega_\mu{}^a{}_c\,\lambda^c  \,.
\end{split}\end{align}
The transformations \eqref{NCtraforules}--\eqref{deltaomega2} together with the constraints \eqref{Rhconstr} and \eqref{nrBianchi}
make up the Newton--Cartan theory of gravity as described in \cite{Andringa:2010it}.

At this point, we may impose equations of motion on the Newton--Cartan gauge-fields, in addition to the constraint
\eqref{Rhconstr} or \eqref{offRhconstr}, for example by performing the limiting procedure on the Einstein
equations. One can show that this leads to the equation of motion presented in \cite{Andringa:2010it}.

%%%%%%%%%%%%%%%%%%%%%%%%%%%%%%%%%%%%%%%%%%%%%%%%%%%%%%%%%%%%%%%%%%%%%%%%%%%%%%%%%%%%%%%%%%%%%%%%%%%%%%%%%%%%%

\subsection{Example 2: three-dimensional on-shell Newton--Cartan
supergravity}\label{subsec:onshellNC}

In this section we will extend the previous example to the
three-dimensional Newton--Cartan supergravity theory constructed in
\cite{Andringa:2013mma}. The reason that we work in three dimensions is that this is the only dimension
in which an example of an on-shell Newton--Cartan supergravity theory is known so far.
Since the discussion mainly parallels the previous
section, we will skip most intermediate steps, where
the contraction parameter $\omega$ is finite, and we will mostly focus on the
results obtained in the $\omega \rightarrow \infty$ limit. Here and in the
following, we will therefore no longer resort to the notation using
tildes, to denote quantities at finite $\omega$.

The underlying gauge algebra in this case is the $\mathcal{N}=2$ Bargmann superalgebra. This superalgebra can be obtained by
contracting the $\mathcal{N}=2$ Poincar\'e superalgebra, with central extension $\mathcal{Z}$, that is given by
\begin{align} \begin{aligned}\label{3dpoincare}
 \big[M_{AB} , \hat P_C \big] &= -2\,\eta_{C[A} \hat P_{B]} \,, &
 \big[M_{AB} , M_{CD} \big] &= 4\,\eta_{[A[C} M_{D]B]} \,, \\[.1truecm]
 \big[M_{AB} , Q^i \big] &= -\frac12\,\gamma_{AB}Q^i \,, &
 \big\{ Q^i , Q^j \big\} &= -\gamma^AC^{-1}\,\hat P_A\,\delta^{ij} +C^{-1}\,\mathcal{Z}\,\epsilon^{ij} \,.
\end{aligned}\end{align}
Here, the supercharges $Q^i\ (i=1,2)$ are two-component
Majorana spinors.
For the gamma-matrices we choose a real basis,
i.e.~$\gamma^A=(i\sigma_2,\sigma_1,\sigma_3)$ and the charge
conjugation matrix is taken to be $C=i\gamma^0$.
In order to define the In\"on\"u--Wigner contraction, we first define
the combinations
\begin{align}\label{Q+-}
 Q_\pm =\frac{1}{\sqrt{2}}\,\big(Q_1 \pm \gamma_0Q_2\big) \,,
\end{align}
and split the three-dimensional flat indices $A$, $B$ into time-like
and space-like indices $\{0,a\}$. As before, we set $M_{ab}=J_{ab}$
for the purely spatial rotations.  The motivation for choosing the
combinations of the relativistic spinors as given in eq.~\eqref{Q+-}, stems from
the non-relativistic algebra (and later on from the transformation
rules of the gravitini). It leads to particularly simple
transformations of the spinors under boosts.  Before making the contraction we perform the following redefinition of the generators:
\begin{align}\begin{aligned}\label{susyalgcontr}
 Q_- &\to \sqrt{\omega}\,Q_- \,, & Q_+&\to \frac{1}{\sqrt\omega}\,Q_+ \,,\hskip.6cm &  M_{a0} &\to \omega \,G_a \,, \\[.1truecm]
 \mathcal{Z} &\to -\omega Z+\frac{1}{2\omega}\,H \,, \hskip.6cm & \hat P_0&\to \omega Z +\frac{1}{2\omega}\,H \,. & &
\end{aligned}\end{align}
Using these redefinitions, the supersymmetric extension
of the Bargmann algebra is then obtained in the limit $\omega\to\infty$, in a similar way as
discussed in the previous subsection. In particular, we find the
following non-vanishing commutation relations:
\begin{align}
 \big[J_{ab} , P_c \big] &= -2\,\delta_{c[a} P_{b]} \,, &
 \big[J_{ab} , G_c \big] &= -2\,\delta_{c[a} G_{b]} \,, \nonumber\\[.12truecm]
 \big[G_a , H \big] &= -P_a \,, & \big[ G_a, P_b \big] &= -\delta_{ab}\,Z \,, \nonumber\\[.12truecm]
 \big[J_{ab} , Q_\pm \big] &= -\tfrac12\,\gamma_{ab}Q_\pm \,, &
 \big[G_a , Q_+ \big] &= -\tfrac12\,\gamma_{a0}Q_- \,,                                           \label{3dsuperbargmann} \\[.12truecm]
 \big\{ Q_+ , Q_+ \big\} &= -\gamma^0C^{-1}\,H \,, &
 \big\{ Q_+ , Q_- \big\} &= -\gamma^aC^{-1}\,P_a \,, \nonumber\\[.12truecm]
 \big\{ Q_- , Q_- \big\} &= -2\,\gamma^0C^{-1}\,Z \,. & & \nonumber
\end{align}
The bosonic part of the algebra corresponds to the Bargmann algebra, see
eq.~\eqref{Bargmannalg}. Note that, since we are working in three dimensions, the
spatial rotations are abelian.

We now wish to extend this contraction to the fields of the on-shell,
relativistic $\mathcal{N}=2$ supergravity multiplet, whose supersymmetry
transformation rules (with parameter $\eta_i$) are given by
\begin{align}
 \delta E_\mu{}^A &= \frac12\,\delta^{ij}\,\bar\eta_i\,\gamma^A \Psi_{\mu j} \,, \label{susyE}\\[.1truecm]
 \delta \Psi_{\mu i} &= D_\mu\eta_i= \partial_\mu\eta_i -\frac14\,\Omega_\mu{}^{AB}(E,\Psi_i)\gamma_{AB}\eta_i \label{susyPsi}\,,
\end{align}
where $D_\mu$ is the Lorentz-covariant derivative and the dependent spin-connection $\Omega_\mu{}^{AB}(E,\Psi_i)$ is given by the supersymmetric analog of \eqref{relomega},
i.e.
 \begin{align}\begin{split}\label{relsusyomega}
 \Omega_\mu{}^{AB}(E,\Psi_i) &= -2\,E^{\rho[A}\Big(\partial_{[\mu} E_{\rho]}{}^{B]}
                                   -\frac14\,\delta^{ij}\,\bar\Psi_{[\mu i}\gamma^{B]}\Psi_{\nu]j}\Big) \\[.1truecm]
   &\quad    +E_{\mu C} E^{\rho A} E^{\nu B} \Big(\partial_{[\rho} E_{\nu]}{}^C
                                   -\frac14\,\delta^{ij}\,\bar\Psi_{[\rho i}\gamma^C\Psi_{\nu]j}\Big)\,.
\end{split}\end{align}
From this expression one derives that the supersymmetry transformation of the (dependent)
spin-connection is given by
\begin{align}\label{susyomegatrafo}
 \delta \Omega_\mu{}^{AB}(E,\Psi_i) = -\frac12\,\delta^{ij}\,E^{\rho[A}\,\bar\eta_i\,\gamma^{B]}\hat\Psi_{\mu\rho j}
         +\frac14\,\delta^{ij}\,E_{\mu C}E^{\rho A}E^{\nu B}\,\bar\eta_i\,\gamma^C\,\hat\Psi_{\rho\nu j}\,.
\end{align}
Note that this transformation rule is zero on-shell, i.e.~it vanishes upon using the fermionic equations of motion
\begin{align}\label{relsusyeom}
 \hat\Psi_{\mu\nu i}=2\,D_{[\mu}\Psi_{\nu]i}=0 \,.
\end{align}
One may verify that the supersymmetry algebra on the fields \eqref{susyE} and \eqref{susyPsi} closes on-shell.

As in the previous subsection, we will introduce a field $M_\mu$, associated
to the central charge transformation $\cal{Z}$ of the $\mathcal{N}=2$ algebra.
Its transformation rule under supersymmetry
is determined by the Poincar\'e superalgebra \eqref{3dpoincare}
\begin{align}\label{susyM}
 \delta M_\mu = \frac12\,\varepsilon^{ij}\,\bar\eta_i\Psi_{\mu j} \,.
\end{align}
This field is ordinarily not introduced in the supergravity
multiplet. In order not to upset the on-shell counting of bosonic and
fermionic degrees of freedom, we are thus obliged to set the
supercovariant curvature of $M_\mu$ to zero, i.e.
\begin{align}\label{susyMcurv}
 \hat F_{\mu\nu}(M) = 2\,\partial_{[\mu}M_{\nu]} -\frac12\,\varepsilon^{ij}\,\bar\Psi_{[\mu i}\Psi_{\nu]j} =0 \,,
\end{align}
so that this field corresponds to a pure gauge degree of freedom. Note that this constraint also implies that the
commutator of two supersymmetry transformations acting on $M_\mu$ closes to a general coordinate transformation
and a central charge transformation. Moreover, since this constraint is the supercovariant version of \eqref{onMcurv},
it will allow us to obtain finite expressions for the
non-relativistic spin-connections from the relativistic one.
Starting from expression (\ref{susyMcurv}), the full set of relativistic equations of motion is obtained by the
following chain of supersymmetry transformations
\begin{align}\label{relchain}
 \hat F_{\mu\nu}(M)=0 \quad\to\quad \hat \Psi_{\mu\nu i}=0  \quad\to\quad \hat R_{\mu\nu}{}^{AB}(\Omega)=0 \,.
\end{align}

This concludes the summary of our relativistic starting point.
Let us now extend the algebra contraction to the fields of this on-shell
supergravity multiplet. For the bosonic fields, this entails the
redefinitions involving $\omega$ that were introduced in the previous
section. The redefinitions of the gravitini follow from the way we
contract the generators of the three-dimensional $\mathcal{N}=2$
Poincar\'e superalgebra to get the Bargmann superalgebra, see the definitions \eqref{Q+-}.
Hence, we define new spinors
\begin{align}\label{psi+-}
 \Psi_\pm =\frac{1}{\sqrt2}\,\Big(\Psi_1 \pm \gamma_0\Psi_2\Big) \,,
\end{align}
and similarly for the parameters $\eta$. We then introduce the scalings:
\begin{align}\begin{split}\label{fermscaling}
 \Psi_{\mu+} &= \sqrt{\omega}\,\psi_{\mu+} \,, \hskip2cm
      \eta_+ = \sqrt{\omega}\,\epsilon_+ \,, \\[.1truecm]
 \Psi_{\mu-} &= \frac{1}{\sqrt{\omega}}\,\psi_{\mu-} \,, \hskip1,9cm
      \eta_- = \frac{1}{\sqrt{\omega}}\,\epsilon_- \,.
\end{split}\end{align}
The following non-relativistic supersymmetry transformation rules then follow
\begin{align}\begin{split}\label{bosfermtrafo}
 \delta \tau_\mu &= \frac12\,\bar\epsilon_+\gamma^0\psi_{\mu+} \,, \\[.1truecm]
 \delta e_\mu{}^a &= \frac12\,\bar\epsilon_+\gamma^a\psi_{\mu-} +\frac12\,\bar\epsilon_-\gamma^a\psi_{\mu+}\,,\\[.1truecm]
 \delta m_\mu &= \bar\epsilon_-\gamma^0\psi_{\mu-} \,,
\end{split}\end{align}
as well as
\begin{align}\begin{split}\label{fermfermtrafo}
 \delta \psi_{\mu+} &= \partial_\mu\epsilon_+  -\frac14\,\omega_\mu{}^{ab}\gamma_{ab}\epsilon_+ \,, \\
 \delta \psi_{\mu-} &= \partial_\mu\epsilon_-  -\frac14\,\omega_\mu{}^{ab}\gamma_{ab}\epsilon_-
                       +\frac12\,\omega_\mu{}^a\gamma_{a0}\epsilon_+ \,.
\end{split}\end{align}
The transformation rules of the spinors with respect to the non-relativistic bosonic symmetries stem
from the relativistic rule $\delta \Psi_\mu =
1/4\,\lambda^{AB}\gamma_{AB}\Psi_\mu$ and are found to be
\begin{align}\begin{split}\label{fermbostrafo}
  \delta \psi_{\mu+} &= \frac14\,\lambda^{ab}\gamma_{ab}\psi_{\mu+} \,, \\
  \delta \psi_{\mu-} &= \frac14\,\lambda^{ab}\gamma_{ab}\psi_{\mu-}
-\frac12\,\lambda^a\gamma_{a0}\psi_{\mu+} \,.
\end{split}\end{align}

It is understood that the spin-connections $\omega_\mu{}^{a},\omega_\mu{}^{ab}$ in \eqref{fermfermtrafo} are dependent,
i.e.~$\omega_\mu{}^{a}=\omega_\mu{}^{a}(e,\tau,m,\psi_\pm)$ and $\omega_\mu{}^{ab}=\omega_\mu{}^{ab}(e,\tau,m,\psi_\pm)$.
The expressions for these non-relativistic spin-connections can be obtained from the relativistic expressions given in eqs.~\eqref{relsusyomega} and \eqref{susyMcurv}. We find
\begin{align}
 \omega_\mu{}^{ab}(e,\tau,m,\psi_\pm) &= -2\,e^{\nu[a}\big(\partial_{[\mu}e_{\nu]}{}^{b]}
                               -\frac12\,\bar\psi_{[\mu+}\gamma^{b]}\psi_{\nu]-}\big)
       +e_\mu{}^ce^{\rho a}e^{\nu b}\big(\partial_{[\rho}e_{\nu]}{}^c
                               -\frac12\,\bar\psi_{[\rho+}\gamma^c\psi_{\nu]-}\big) \nonumber\\
    &\quad  -\tau_\mu e^{\rho a}e^{\nu b}\big(\partial_{[\rho}m_{\nu]} -\frac12\,\bar\psi_{[\rho-}\gamma^0\psi_{\nu]-} \big)
            \,, \label{susyomegaab} \\
\begin{split}\label{susyomegaa}
 \omega_\mu{}^a(e,\tau,m,\psi_\pm) &= \tau^\nu \big(\partial_{[\mu}e_{\nu]}{}^a
                               -\frac12\,\bar\psi_{[\mu+}\gamma^a\psi_{\nu]-} \big)
      +e_{\mu b} e^{\rho a}\tau^\nu \big(\partial_{[\rho}e_{\nu]}{}^b -\frac12\,\bar\psi_{[\rho+}\gamma^b\psi_{\nu]-} \big) \\
    &\quad  +e^{\nu a}\big(\partial_{[\mu}m_{\nu]} -\frac12\,\bar\psi_{[\mu-}\gamma^0\psi_{\nu]-} \big)
            -\tau_\mu e^{\rho a}\tau^\nu \big(\partial_{[\rho}m_{\nu]}
                                      -\frac12\,\bar\psi_{[\rho-}\gamma^0\psi_{\nu]-} \big) \,.
\end{split}
\end{align}
In order to obtain these expression, we have mimicked the discussion around equations
\eqref{omegaabdef}--\eqref{Mcurvexpl}. This time however, we have used
eq.~\eqref{susyMcurv} to replace terms that diverge in the $\omega \to \infty$
limit, by terms with the expected $\omega$-order. Like in the bosonic case, the above expressions for the spin-connections
identically solve the supercovariant curvature constraints
\begin{align}\begin{split}\label{RPRZ}
 \hat R_{\mu\nu}{}^a(P) & = R_{\mu\nu}{}^a(P) -\bar\psi_{[\mu+}\gamma^a\psi_{\nu]-}=0 \,, \\[.1truecm]
 \hat R_{\mu\nu}(Z) & = R_{\mu\nu}(Z) -\bar\psi_{[\mu-}\gamma^0\psi_{\nu]-}=0\,.
\end{split}\end{align}
The so-called conventional constraints \eqref{RPRZ} are identically fulfilled,
so we need not worry about
variations thereof. They can be used to determine the transformations of the
spin- and boost-connections
\eqref{susyomegaab}
and \eqref{susyomegaa}.

The $\omega \to \infty$ limit of \eqref{susyMcurv} leads to the further constraint
\begin{align}\label{susyRhcurv}
 \hat R_{\mu\nu}(H) = R_{\mu\nu}(H)
-\frac12\,\bar\psi_{[\mu+}\gamma^0\psi_{\nu]+} =0\,.
\end{align}
This leads to further conditions upon variation under supersymmetry. To check this
we need to know the supersymmetry variations of the spin-connections. Using
the transformation rules \eqref{bosfermtrafo} and
\eqref{fermfermtrafo} in the expressions \eqref{susyomegaab}, \eqref{susyomegaa} we find
\begin{align}
\begin{split}\label{susyomegaabtrafo}
 \delta_Q \omega_\mu{}^{ab}(e,\tau,m,\psi_\pm) &= \frac12\,\bar\epsilon_+\gamma^{[b}\hat\psi^{a]}{}_{\mu-}
         +\frac14\,e_{\mu c} \,\bar\epsilon_+\gamma^c\hat\psi^{ab}{}_-
         -\frac12\,\tau_\mu \,\bar\epsilon_-\gamma^0\hat\psi^{ab}{}_- \\[.1truecm]
   &\quad +\frac12\,\bar\epsilon_-\gamma^{[b}\hat\psi^{a]}{}_{\mu+}
         +\frac14\,e_{\mu c} \,\bar\epsilon_-\gamma^c\hat\psi^{ab}{}_+\,,
\end{split} \\[.1truecm]
\begin{split}\label{susyomegaatrafo}
  \delta_Q \omega_\mu{}^a(e,\tau,m,\psi_\pm) &= \frac12\,\bar\epsilon_-\gamma^0\hat\psi_\mu{}^a{}_-
         +\frac12\,\tau_\mu\,\bar\epsilon_-\gamma^0\hat\psi_0{}^a{}_-
         +\frac14\,e_{\mu b}\,\bar\epsilon_+\gamma^b\hat\psi^a{}_{0-}
         +\frac14\,\bar\epsilon_+\gamma^a\hat\psi_{\mu0-}\\[.1truecm]
   &\quad +\frac14\,e_{\mu b}\,\bar\epsilon_-\gamma^b\hat\psi^a{}_{0+}
         +\frac14\,\bar\epsilon_-\gamma^a\hat\psi_{\mu0+} \,.
\end{split}
\end{align}
Now we readily derive that under supersymmetry transformations, with parameters $\epsilon_+$ and  $\epsilon_-$, the
following set of constraints is generated:
\begin{align}
 \stackrel{Q_-}{\longrightarrow} \hskip1.1cm \hat\psi_{ab-} &=0 \label{moreconstraints}\\
 \hat R_{\mu\nu}(H)=0 \quad \stackrel{Q_+}{\longrightarrow} \quad \hat\psi_{\mu\nu+}=0 \quad
    \stackrel{Q_+}{\longrightarrow} \quad 
             R_{\mu\nu}{}^{ab}(J) &=0 \label{constraints} \\
  \stackrel{Q_+}{\longrightarrow} \hskip.7cm \gamma^a\hat\psi_{a0-} &=0
             \quad\stackrel{Q_+}{\longrightarrow}\quad \hat R_{0a}{}^a(G)=0 \label{onshell}
\end{align}
Note that the  variation of the $\hat\psi_{\mu\nu+}=0$ constraint leads to three different constraints. Of
these three constraints only the variation of the
$ \gamma^a\hat\psi_{a0-}=0$ leads to one more constraint.
Using the last constraint  given in eq.~\eqref{constraints} the non-relativistic Bianchi identities reduce to
\begin{align}\label{susynrBianchi}
 \hat R_{ab}{}^c(G)=0 \,, \hskip2cm \hat R_{0[a}{}^{b]}(G) =0 \,.
\end{align}
These identities are e.g.~needed to show that the variation of the constraint given in eq.~\eqref{moreconstraints}
does not lead to further constraints.
We did not check the  variation of the last constraint in eq.~\eqref{onshell}. The calculation  is quite
involved and has also not been carried out in
\cite{Andringa:2013mma}. We will, however, show in section
\ref{sec:offshellsugra} that the full set of constraints
\eqref{moreconstraints}--\eqref{onshell} can be derived from an
off-shell version of this multiplet where we have checked the
consistency of the whole set of constraints.

At this point we have finished the derivation of the three-dimensional
on-shell Newton--Cartan supergravity constructed in
\cite{Andringa:2013mma}, i.e.~we obtained all constraints and
transformation rules. The terminology `on-shell' stems from the fact
that the constraints given in eq.~(\ref{onshell}) both can be interpreted as equations of motion for Newton--Cartan
supergravity:  the first constraint is
necessary to obtain closure of the supersymmetry algebra while the bosonic part of the second constraint
is precisely the equation of motion of the bosonic Newton--Cartan gravity theory.
Note, however, that to call some constraints `equation of motion' and others not is
slightly ambiguous when talking about Newton--Cartan (super)gravity,
due to the absence of an action principle that can be used to derive
these equations of motion. In section \ref{sec:offshellsugra}, we will
construct a different, `off-shell' version of three-dimensional
Newton--Cartan supergravity, that includes an auxiliary scalar field
in the supermultiplet. The terminology `off-shell' will be justified
in the sense that the first constraint given in eq.~(\ref{onshell}) will no
longer be needed for closure of the supersymmetry algebra. Both
constraints given in eq.~(\ref{onshell}) will in fact not appear at
all. Equations of motion can thus be identified in a pragmatic way as
those constraints that can be removed by adding auxiliary degrees of
freedom to a non-relativistic supermultiplet.

Let us stress/repeat some important aspects of this second example. In the case at hand we
can draw the commuting diagram, given in figure \ref{fig:1}. The left column represents a chain of
relativistic constraints while the right column contains a similar chain of non-relativistic constraints.
\begin{figure}[h!]
\begin{center}
\begin{tikzpicture}
    \draw [black, thick, ->] (-1,2) -- (-1,1.3) ;
    \draw [black, thick, ->] (-1,.5) -- (-1,-.2) ;

    \node [black,above] at (-.6,1.3) {\footnotesize $Q_i$} ;
    \node [black,above] at (-.6,-.2) {\footnotesize $Q_i$} ;

    \node [black,above] at (-1,2) {$\hat F_{\mu\nu}(M)=0$} ;
    \node [black,above] at (-1,.5) {$\hat \Psi_{\mu\nu i}=0$} ;
    \node [black,above] at (-1,-1) {$ R_{\mu\nu}{}^{AB}(\Omega)=0$} ;

    \draw [black, thick, ->] (8,2) -- (8,1.3) ;
    \draw [black, thick, ->] (8,.5) -- (8,-.2) ;

    \node [black,above] at (8.4,1.3) {\footnotesize $Q_+$} ;
    \node [black,above] at (8.4,-.2) {\footnotesize $Q_+$} ;

    \node [black,above] at (8,2) {$\hat R_{\mu\nu}(H)=0$} ;
    \node [black,above] at (8,.5) {$\hat \psi_{\mu\nu+}=0$} ;
    \node [black,above] at (8,-1) {$ R_{\mu\nu}{}^{ab}(J)=0$} ;

  \draw [black, thick, ->] (1.6,2.3)--(5.4,2.3) ;
    \draw [black, thick, ->] (1.6,.8)--(5.4,.8) ;
       \draw [black, thick, ->] (1.6,-.7)--(5.4,-.7) ;
 %   \draw [black, thick, ->] (0.8,2.3)--(6.2,2.3) ;
  %  \draw [black, thick, ->] (0.8,.8)--(6.2,.8) ;
   % \draw [black, thick, ->] (0.8,-.7)--(6.2,-.7) ;
    \node [black,above] at (3.5,2.4) {\footnotesize $\omega\to\infty$} ;
    \node [black,above] at (3.5,.9) {\footnotesize $\omega\to\infty$} ;
    \node [black,above] at (3.5,-.6) {\footnotesize $\omega\to\infty$} ;

\end{tikzpicture}
      \caption{The chains of constraints in the relativistic (left column) and non-relativistic (right column)
case. In the non-relativistic case we do
not denote the complete chain, as given in  eqs.~\eqref{moreconstraints}--\eqref{onshell}.}
      \label{fig:1}
\end{center}
\end{figure}
The diagram shows all non-relativistic constraints that are obtained by a $\omega\to\infty$ limit of the  relativistic ones.
However, in this way we do not obtain the full set of non-relativistic constraints.
This is due to the fact that in the left column we have included both supersymmetries
but in the right column we have only included the variation under $Q_+$ transformations. Further
constraints follow from the variation under the $Q_-$ transformations, but
those non-relativistic constraints are not obtained as  limits of relativistic constraints.

We note that  in the
limit $\omega\to\infty$ the two relativistic constraints given in the second row of the left column, namely those containing the
two gravitino curvatures,
lead to just the single  non-relativistic constraint given in the second row of the right column.
 This is in
line with the fact that the constraint $\hat R_{\mu\nu}(H)=0$ only
varies under one of the two non-relativistic supersymmetries and hence
its variation under supersymmetry  only leads to one of the non-relativistic gravitino curvatures. This
observation is of vital importance to understand the off-shell case
treated in section \ref{sec:offshellsugra}.  There we are also going
to impose the constraint $\hat F_{\mu\nu}(M)=0$, but since its
non-relativistic limit does not necessarily lead to the
non-relativistic equations of motion, imposing this constraint does
not force us to go to the non-relativistic on-shell multiplet of the
current section.

We finish this section with a third illustration  of the
non-relativistic limiting procedure in which we consider a superparticle in a curved background.

%%%%%%%%%%%%%%%%%%%%%%%%%%%%%%%%%%%%%%%%%%%%%%%%%%%%%%%%%%%%%%%%%%%%%%%%%%%%%%%%%%%%%%%%%%%%%%%%%%%%%%%%%%%%%

\subsection{Example 3: the non-relativistic  superparticle in a curved background}\label{subsec:curvpp}

In this third example we apply the limiting procedure to a superparticle moving in a curved background.
To be concrete,  we use it to derive the action and transformation rules of the non-relativistic
superparticle in a curved background,
put forward in \cite{Bergshoeff:2014gja}. The non-relativistic
superparticle in a flat background was already discussed in
\cite{Casalbuoni:1976tz,Brink:1981nb,Siegel:1983hh}. We note that the limit that was taken
in \cite{Gomis:2004pw} to derive the non-relativistic superparticle in a flat background can be
understood as a special case of the analysis in this section.

It is illustrative to first discuss the bosonic particle. To derive the action of a non-relativistic
bosonic point-particle in an arbitrary Newton--Cartan
background we start from the relativistic action
\begin{align}\label{relpp}
 S_{\rm rel}=-M\int\!d\lambda\,\Big( \sqrt{-\eta_{AB}(\dot x^\mu E_\mu{}^A)(\dot x^\nu E_\nu{}^B)} -\dot x^\mu M_\mu\Big)\,.
\end{align}
All dots refer to derivatives w.r.t.~the worldline parameter $\lambda$, i.e.~$\dot x^\mu=dx^\mu/d\lambda$. We use mostly
plus signature and we also added a ``charge'' term $\dot x^\mu M_\mu$. Here, we impose that the curvature of the
abelian gauge-field $M_\mu$ vanishes, implying that it can locally be written as $M_\mu=\partial_\mu \Gamma$ and the second
term in \eqref{relpp} corresponds to a total derivative. Using the expressions \eqref{Eexpand} and
\eqref{Mlimit} in the relativistic action \eqref{relpp} and taking $M=\omega\, m$, we obtain, in the
limit $\omega\rightarrow\infty$, the following non-relativistic action:
\begin{align}\label{nrpp}
 S_{\rm nr}= m\int\!d\lambda\,\bigg[\frac{\delta_{ab}(\dot x^\mu e_\mu{}^a)(\dot x^\nu e_\nu{}^b)}{2\tau_\rho\dot x^\rho}
                                     -m_\mu \dot x^\mu \bigg] \,.
\end{align}
This action
agrees with the action, calculated by other means, in e.g.~\cite{Kuchar:1980tw,DePietri:1994je,Andringa:2012uz}. Note that one of the
reasons to add the term $\dot x^\mu M_\mu$ is to cancel a divergent (total derivative)
term that otherwise would arise in the limit $\omega\rightarrow\infty$, see also \cite{Gomis:2000bd}. In contrast, the
combination $\dot x^\mu m_\mu$
in the non-relativistic action is not
a total derivative term. This non-relativistic term does not only follow from the relativistic $\dot x^\mu M_\mu$ term, but it also
receives contributions from the kinematic term $\sqrt{-\dot x^2}$.

We  now generalize the
discussion of the non-relativistic bosonic particle to the non-relativistic superparticle.
The relativistic superparticle in a curved background is most conveniently written using superspace techniques,
see \cite{Witten:1985nt}. Since so far a non-relativistic superspace description is lacking, we will refrain from using
superspace notation and simplify the
discussion and notation by considering only the terms in the action that are at most quadratic in the fermions.
Thus, the supersymmetric analog of \eqref{relpp} takes the
form
\begin{align}\label{relcurvpp}
 S_{\rm rel}= -M\int\!d\lambda\,\Big[\sqrt{-\eta_{AB}\,\Pi^A\Pi^B}
                      -\frac14\,\varepsilon^{ij}\,\bar\theta_i D_\lambda\theta_j
                      -\dot x^\mu\,\big(M_\mu -\frac12\,\varepsilon^{ij}\,\bar\theta_i\,\Psi_{\mu j}\big)\Big] \,.
\end{align}
The background fields $E_\mu{}^A$, $M_\mu$ and $\Psi_{\mu i}$ are those of the three-dimensional on-shell theory
discussed in section \ref{subsec:onshellNC} and the embedding coordinates are $x^\mu$ and $\theta_i$.
The supersymmetric line-element $\Pi^A$ is defined as
\begin{align}
 \Pi^A = \dot x^\mu\big(E_\mu{}^A-\frac12\,\delta^{ij}\,\bar\theta_i\,\gamma^A\Psi_{\mu j} \big)
                   +\frac14\,\delta^{ij}\,\bar\theta_i\,\gamma^A D_\lambda\theta_j \,,
\end{align}
where the derivative $D_\lambda$ is covariantized w.r.t.~Lorentz transformations, i.e.
\begin{align}
 D_\lambda\theta = \dot\theta-\frac14\,\dot x^\mu\,\Omega_\mu{}^{AB}(E,\Psi_i)\,\gamma_{AB}\theta \,.
\end{align}
As we are only interested in terms up to second order in fermions expressions like $\eta_{AB}\Pi^A\Pi^B$ are
understood to contain only such terms and all terms quartic in fermions are  discarded.

The action \eqref{relcurvpp} is invariant under the following supersymmetry transformations of the embedding
coordinates
\begin{align}
 \delta x^\mu = -\frac14\,\delta^{ij}\,\bar\eta_i\,\gamma^A\theta_j\,E^\mu{}_A \,, \hskip2cm
 \delta \theta_i = \eta_i \,.
\end{align}
These transformations should be accompanied by the following $\sigma$-model transformations \cite{Moore:1984dc,Hull:1995gk}
of the background fields,
 as explained e.g.~in \cite{Andringa:2012uz,Bergshoeff:2014gja}:
\begin{align}\begin{split}
 \delta E_\mu{}^A &= \frac12\,\delta^{ij}\,\bar\eta_i\,\gamma^A\Psi_{\mu j}
              -\frac14\,\delta^{ij}\,\bar\eta_i\,\gamma^B\theta_j\,E^\rho{}_B\partial_\rho E_\mu{}^A \,,
 \hskip1,5cm \delta \Psi_{\mu i} = D_\mu \eta_i \,, \\[.1truecm]
 \delta M_\mu &= \frac12\,\varepsilon^{ij}\,\bar\eta_i\,\Psi_{\mu j}
              -\frac14\,\delta^{ij}\,\bar\eta_i\,\gamma^B\theta_j\,E^\rho{}_B\partial_\rho M_\mu \,.
\end{split}\end{align}
The action \eqref{relcurvpp} is also left invariant by the $\kappa$-transformations
\begin{align}\label{coordkappa}
 \delta_\kappa x^\mu = -\frac14\,\delta^{ij}\,\bar\theta_i\,\gamma^A\delta_\kappa\theta_j\,E^\mu{}_A \,, \hskip1,5cm
 \delta_\kappa \theta_1 = \kappa \,, \hskip1cm \delta_\kappa\theta_2= -\frac{\Pi^A\gamma_A}{\sqrt{-\Pi^2}}\,\kappa \,.
\end{align}
In this case all  background fields transform under $\kappa$-symmetry only through their dependence on the embedding coordinates.
To show invariance under supersymmetry and
$\kappa$-symmetry, one needs to use the equations of motion
of the background fields.

With all these preliminaries at hand it is now straightforward to apply the limiting procedure to the relativistic
superparticle action \eqref{relcurvpp}. This yields the following result:
\begin{align}\begin{split}\label{nrcurvpp}
 S_{\rm nr} &= \frac{m}{2}\int\!d\lambda\,\bigg[
    \frac{\hat\pi^a\hat\pi^b\delta_{ab}}{\hat\pi^0}-2\,\dot x^\mu \big(m_\mu-\bar\theta_-\gamma^0\psi_{\mu-}\big)
    -\bar\theta_-\gamma^0\hat D\theta_-
      -\frac12\,\dot x^\mu\omega_\mu{}^a\,\bar\theta_+\gamma_a\theta_- \bigg] \,,
\end{split}\end{align}
where we have defined the following supersymmetric line elements
\begin{align}
 \hat\pi^0 &=\dot x^\mu\big(\tau_\mu-\frac12\,\bar\theta_+\gamma^0\psi_{\mu+}\big)
             +\frac14\,\bar\theta_+\gamma^0\hat D\theta_+ \,, \\[.1truecm]
\begin{split}\label{piahat}
 \hat\pi^a &=\dot x^\mu\big(e_\mu{}^a-\frac12\,\bar\theta_+\gamma^a\psi_{\mu-}-\frac12\,\bar\theta_-\gamma^a\psi_{\mu+}\big)
           +\frac14\,\bar\theta_+\gamma^a\hat D\theta_- +\frac14\,\bar\theta_-\gamma^a\hat D\theta_+ \\[.1truecm]
         &\qquad  +\frac18\,\bar\theta_+\gamma^a\gamma_{b0}\theta_+\,\dot x^\mu\omega_\mu{}^b \,.
\end{split}
\end{align}
Note that the supercovariant derivative $\hat D$ is covariant  w.r.t.~spatial rotations, not boosts. The boost-connection $\omega_\mu{}^a$ that
appears in eqs.~\eqref{nrcurvpp} and \eqref{piahat} is the dependent boost-con\-nec\-tion \eqref{susyomegaa}. For notational simplicity we
do not denote below its dependence on the other fields. The transformations of the embedding coordinates under $\kappa$-symmetry are given by
\begin{align}\begin{split} \label{nrkappa}
 \delta t &= -\frac14\,\bar\theta_+\gamma^0\kappa \,, \hskip4,78cm \delta \theta_+ = \kappa\,, \\
 \delta x^i &= -\frac14\,\bar\theta_-\gamma^i\kappa -\frac18\frac{\hat\pi^k}{\hat\pi^0}\,\bar\theta_+\gamma^{0i}\gamma_k\kappa\,,
             \hskip2cm   \delta \theta_- = -\frac{\hat\pi^i}{2\hat\pi^0}\,\gamma_{i0}\kappa   \,.
\end{split}\end{align}
This reproduces precisely, to second order in fermions,  the $\kappa$-symmetric
non-relativistic superparticle in a
curved background as presented in \cite{Bergshoeff:2014gja}.

Fixing kappa-symmetry by setting $\theta_+=0$
we  obtain the result of \cite{Bergshoeff:2014gja}.
When we gauge-fix the Newton--Cartan background to a Galilean background with a Newton potential $\Phi$,
the one described in \cite{Andringa:2013mma},
the action \eqref{nrcurvpp} reduces to
\begin{align}\label{nrgalpp}
 S_{\rm nr} &= \frac{m}{2}\int\!d\lambda\,\Big[
    \frac{\pi_\Phi^i\pi_\Phi^i}{\pi^0}-2\dot t \big(\Phi-\bar\theta_-\gamma^0\Psi\big)
  -\bar\theta_-\gamma^0\dot\theta_-
      +\frac{\dot t}{2}\,\partial_i\Phi\,\bar\theta_+\gamma^i\theta_- \Big] \,,
\end{align}
with the `super-Galilean' line-elements given by
\begin{align}\label{phipi}
      \pi^0 &=\dot t +\frac14\,\bar\theta_+\gamma^0\dot\theta_+ \,, \\[.1truecm]
 \pi_\Phi^i &=\dot x^i -\frac12\,\dot t\,\bar\theta_+\gamma^i\Psi +\frac14\,\bar\theta_+\gamma^i\dot\theta_-
           +\frac14\,\bar\theta_-\gamma^i\dot\theta_+
           -\frac18\,\dot t\,\partial_j\Phi\,\bar\theta_+\gamma^i\gamma_{j0}\theta_+ \,.
\end{align}
This finishes our discussion of the superparticle in a non-relativistic curved background.

%%%%%%%%%%%%%%%%%%%%%%%%%%%%%%%%%%%%%%%%%%%%%%%%%%%%%%%%%%%%%%%%%%%%%%%%%%%%%%%%%%%%%%%%%%%%%%%%%%%%%%%%%%%%%
%%%%%%%%%%%%%%%%%%%%%%%%%%%%%%%%%%%%%%%%%%%%%%%%%%%%%%%%%%%%%%%%%%%%%%%%%%%%%%%%%%%%%%%%%%%%%%%%%%%%%%%%%%%%%

\section{3D non-relativistic off-shell supergravity}\label{sec:offshellsugra}

In this section, we apply the limiting procedure to obtain an \emph{off-shell} version
of the Newton--Cartan supergravity theory of \cite{Andringa:2013mma}, that was
revisited in section \ref{subsec:onshellNC}. Such an off-shell version will
necessarily contain auxiliary fields, that cannot be interpreted as gauge-fields
of an underlying symmetry algebra. One can therefore not use a gauging
procedure to obtain this theory. This example shows
that the limiting procedure provides us with an effective tool to obtain
non-relativistic theories from relativistic ones, when no gauging procedure is
available.

We will start from an off-shell formulation of three-dimensional $\mathcal{N}=2$
supergravity. There exist two different such off-shell formulations \cite{
Achucarro:1987vz,Achucarro:1989gm,Howe:1995zm}. We will start from the
so-called three-dimensional new minimal $\mathcal{N}=2$ Poincar\'e
multiplet, since this multiplet contains the abelian central charge gauge-field
$M_\mu$ that was already needed in the on-shell case.
Due to the lack of a vector gauge-field with the transformation rule
given by eq.~\eqref{susyM} it is not obvious how to take the non-relativistic limit of the old minimal
$\mathcal{N}=2$ Poincar\'e multiplet.

The new minimal  multiplet
consists of a dreibein $E_\mu{}^A$, two gravitini $\Psi_{\mu i}\ (i=1,2)$,
two auxiliary vector gauge-fields $M_\mu$ and $V_\mu$ and an auxiliary scalar $D$, see e.g.~\cite{Howe:1995zm}.
The supersymmetries (with parameters $\eta_i$), the central charge transformations (with parameter $\Lambda$)
and the R-symmetry transformations (with parameter $\rho$) of these fields are given by
\begin{align}\begin{split}\label{offshelltrafos}
 \delta E_\mu{}^A &= \frac12\,\delta^{ij}\,\bar\eta_i\,\gamma^A\,\Psi_{\mu j} \,, \\[.1truecm]
 \delta \Psi_{\mu i} &= D_\mu\eta_i +\varepsilon^{ij}\,\eta_j\,V_\mu -\gamma_\mu\eta_i\,D
                     +\frac14\,\gamma_\mu\,\gamma\cdot\hat F(M)\,\varepsilon^{ij}\,\eta_j
                     -\varepsilon^{ij}\,\Psi_{\mu j}\,\rho \,, \\[.1truecm]
 \delta M_\mu &= \frac12\,\varepsilon^{ij}\,\bar\eta_i\, \Psi_{\mu j} + \partial_\mu\Lambda \,, \\[.1truecm]
 \delta V_\mu &= \frac12\,\varepsilon^{ij}\,\bar\eta_i\, \gamma^\nu \hat\Psi_{\mu\nu j}
                     -\frac18\,\varepsilon^{ij}\,\bar\eta_i\,\gamma_\mu\gamma\cdot\hat\Psi_j
                     -\frac14\,\delta^{ij}\,\bar\eta_i\,\gamma\cdot\hat F(M)\, \Psi_{\mu j}
                     -\varepsilon^{ij}\,\bar\eta_i\, \Psi_{\mu j}\,D +\partial_\mu\rho\,, \\[.1truecm]
 \delta D &= -\frac{1}{16}\,\delta^{ij}\,\bar\eta_i\, \gamma\cdot\hat\Psi_j \,.
\end{split}\end{align}
The field strength $\hat F_{\mu\nu}(M)$ of the central charge gauge field is given by eq.~\eqref{susyMcurv} while the two gravitino
curvatures read
\begin{align}
 \hat \Psi_{\mu\nu i} = 2\,D_{[\mu}\Psi_{\nu]i} -2\,\gamma_{[\mu}\Psi_{\nu]i}\,D
                     -2\,\varepsilon^{ij}\,\Psi_{[\mu j}\,V_{\nu]}
                     +\frac12\,\varepsilon^{ij}\,\gamma_{[\mu}\gamma\cdot\hat F(M)\,\Psi_{\nu]j} \,.
\end{align}
The dots refer to gamma traces as in $\gamma\cdot\hat F(M)=\gamma^{\mu\nu}\hat F_{\mu\nu}(M)$. The spin-connection is
determined by requiring that the supercovariant torsion $\hat R_{\mu\nu}{}^A(E)$ is zero. Its supersymmetry variation follows from
the expression in terms of $E_\mu{}^A$ and $\Psi_{\mu i}$, see eq.~\eqref{relsusyomega}.

In order to apply the limiting procedure to the transformation rules given in eq.~\eqref{offshelltrafos}, we use
the same rescalings as in the previous sections, supplemented with
\begin{align}
 D\to \frac1\omega\,S \,.
\end{align}
We do not rescale the auxiliary field $V_\mu$. Below we will argue
that in the non-relativistic limit one must eliminate $V_\mu$.
The action of $\mathcal{N}=2$ new minimal supergravity contains a
$D^2$ term that plays the role of the cosmological
constant $\Lambda_{CC}$. It is thus not surprising that in the non-relativistic
limit, $D$ scales like the square root of
$\Lambda_{CC}$, see e.g.~\cite{Bergshoeff:2014gja} for the non-relativistic
contraction of the anti de Sitter algebra.

Going through similar arguments as in sections \ref{subsec:NC} and
\ref{subsec:onshellNC} we determine the non-relativistic spin-connections
$\omega_\mu{}^{ab}(e,\tau,m,\psi_\pm)$ and $\omega_\mu{}^a(e,\tau,m,\psi_\pm)$
to be given by eqs.~\eqref{susyomegaab} and \eqref{susyomegaa}. As in the on-shell
case, we need to impose eq.~\eqref{susyMcurv} as an extra constraint, in order to
take the non-relativistic limit consistently. In fact, to get the correct expressions for the spin-connections
we only need to set to zero the spatial components of $\hat F_{\mu\nu}(M)$. However, this is not sufficient to
take the non-relativistic limit in the transformation rules of all fields. Indeed, we need to eliminate also
all remaining components of $\hat F_{\mu\nu}(M)$ as well as of  $\hat\psi_{\mu\nu+}$ to avoid divergent terms in the transformation
rules of the fields.
Since the relativistic constraint $\hat F_{\mu\nu}(M)=0$ varies under supersymmetry to the fermionic equations of motion,
we effectively
put the relativistic theory on-shell.
In the following, we will show that, upon elimination of only one of the auxiliary fields, the limiting procedure leads to an
irreducible non-relativistic multiplet on which the Bargmann superalgebra is realized off-shell, in a sense that we will
clarify below.

Here, we present a brief discussion to argue why we can
eliminate the auxiliary field $V_\mu$. In a first approach the
limiting procedure leads to the constraints
\begin{align}\label{offshellconstr1}
 \hat R_{\mu\nu}(H)=0 \,, \hskip2cm \hat\psi_{\mu\nu+}=0 \,, \hskip2cm \hat\psi_{ab-}=0 \,.
\end{align}
At this point we derive the following transformation rules for $\tau_\mu$ and the auxiliary fields $V_\mu$ and $S$:
\begin{align}\begin{split}
 \delta \tau_\mu &= \frac12\,\bar\epsilon_+\gamma^0\psi_{\mu+} \,, \\
 \delta V_\mu &= -\frac14\,\bar\epsilon_+\gamma^{a0}\hat\psi_{\mu a-} -\bar\epsilon_+\gamma^0\psi_{\mu+}\,S \,, \\[.1truecm]
 \delta S &= -\frac18\,\bar\epsilon_+\gamma^{a0}\hat\psi_{a0-} \,.
\end{split}\end{align}
The supersymmetry transformations of the last two constraints of \eqref{offshellconstr1} imply
\begin{align}\label{Vconstr}
 e^\mu{}_a e^\nu{}_b \,\hat V_{\mu\nu}= \hat V_{ab}=0 \,,
\end{align}
which is the spatial part of the supercovariant curvature of $V_\mu$:
\begin{align}\label{Vcurv}
 \hat V_{\mu\nu} = 2\,\partial_{[\mu}V_{\nu]} +\frac12\,\bar\psi_{[\mu+}\gamma^{a0}\hat\psi_{\nu]a-}
                  +\bar\psi_{[\mu+}\gamma^0\psi_{\nu]+}\,S \,.
\end{align}
Using the first and last constraint in eq.~\eqref{offshellconstr1} we observe that the constraint \eqref{Vconstr}
is always satisfied if we set
\begin{align}\label{noV}
 V_\mu = -2\,\tau_\mu\,S \,.
\end{align}
The inverse vielbeins in \eqref{Vconstr} eliminate any term with a
free $\tau_\mu$ and thus the derivative in \eqref{Vcurv} must hit the
$\tau_\mu$ when we insert eq.~\eqref{noV} in expression \eqref{Vcurv}. We can then
use the first constraint of \eqref{offshellconstr1} to cancel all
remaining terms.  Furthermore, the identification \eqref{noV} is
preserved under all symmetry transformations, upon use of the constraints given in eq.~\eqref{offshellconstr1}.
In particular, the combination
$V_\mu+2\,\tau_\mu\,S$ does not transform under supersymmetry. It is
therefore not needed to realize the supersymmetry algebra
off-shell. With the aim of deriving an irreducible multiplet we shall
therefore eliminate $V_\mu$, using \eqref{noV}. This sets the
R-symmetry parameter $\rho={\rm const}$ in \eqref{offshelltrafos}.

Performing the above manipulations, we end up with the following transformation rules for the complete non-relativistic
off-shell multiplet
\begin{align}\begin{split}
 \delta \tau_\mu &= \frac12\,\bar\epsilon_+\gamma^0\psi_{\mu+} \,, \\
 \delta e_\mu{}^a &= \frac12\,\bar\epsilon_+\gamma^a\psi_{\mu-} +\frac12\,\bar\epsilon_-\gamma^a\psi_{\mu+}\,,\\
 \delta m_\mu &= \bar\epsilon_-\gamma^0\psi_{\mu-} \,, \\[.2truecm]
 \delta \psi_{\mu+} &= D_\mu\epsilon_+ +\gamma_0\epsilon_+\,S\,\tau_\mu +\gamma_0\psi_{\mu+}\,\rho \,, \\
 \delta \psi_{\mu-} &= D_\mu\epsilon_- -3\,\gamma_0\epsilon_-\,S\,\tau_\mu
                +\frac12\,\omega_\mu{}^a\gamma_{a0}\epsilon_+ -\gamma_a\epsilon_+\,e_\mu{}^a\,S
                -\gamma_0\psi_{\mu-}\,\rho\,, \\
 \delta S &= -\frac18\,\bar\epsilon_+\gamma^{a0}\hat\psi_{a0-} \,.
\end{split}\end{align}
Given that in the non-relativistic case there is only a single
(fermionic) equation of motion in the on-shell theory, see
eq.~\eqref{onshell}, it is not surprising that the number of auxiliary
fields, needed to close the algebra off-shell, is reduced with respect
to the relativistic multiplet we started with.

We have explicitly checked that the non-relativistic supersymmetry
transformations given above close off-shell, i.e.~upon use of the
constraints \eqref{offmoreconstraints}--\eqref{offBianchi} given below. Note that
the commutator algebra closes off-shell in the sense that
we do not need the equations of motion \eqref{onshell} to prove closure.
To check closure one needs the supersymmetry transformations of the
spin- and boost-connection
\begin{align}
 \delta_Q \omega_\mu{}^{ab}(e,\tau,m,\psi_\pm) &= \frac12\,\bar\epsilon_+\,\gamma^{[b}\hat\psi^{a]}{}_{\mu-}
                -S\,\bar\epsilon_+\gamma^{ab}\psi_{\mu+} \,, \\[.1truecm]
\begin{split}
 \delta_Q \omega_\mu{}^a(e,\tau,m,\psi_\pm) &= \frac14\,\bar\epsilon_+\gamma^a\hat\psi_{\mu0-}
                +\frac14\,e_{\mu b}\,\bar\epsilon_+\gamma^b\hat\psi^a{}_{0-} +S\,\bar\epsilon_+\gamma^{a0}\psi_{\mu-} \\[.1truecm]
        &\quad  +\bar\epsilon_-\gamma^0\hat\psi_\mu{}^a{}_- +S\,\bar\epsilon_-\gamma^{a0}\psi_{\mu+} \,,
\end{split}
\end{align}
as well as the expressions for the gravitino curvatures
\begin{align}\begin{split}
 \hat \psi_{\mu\nu+} &= 2\,\partial_{[\mu}\psi_{\nu]+}-\frac12\,\omega_{[\mu}{}^{ab}\gamma_{ab}\psi_{\nu]+}
              -2\,\gamma_0\psi_{[\mu+}\,\tau_{\nu]} \,S \,, \\[.1truecm]
 \hat \psi_{\mu\nu-} &= 2\,\partial_{[\mu}\psi_{\nu]-}-\frac12\,\omega_{[\mu}{}^{ab}\gamma_{ab}\psi_{\nu]-}
              +6\,\gamma_0\psi_{[\mu-}\,\tau_{\nu]} \,S +\omega_{[\mu}{}^a\gamma_{a0}\psi_{\nu]+}
              +2\,\gamma_a\psi_{[\mu+}\,e_{\nu]}{}^a\,S \,.
\end{split}\end{align}
The commutator of two supersymmetry transformations is given by
\begin{align}
 \big[\delta_Q(\epsilon_1),\delta_Q(\epsilon_2)\big] =
\delta_{g.c.t.}\big(\Xi^\rho\big) +\delta_J\big(\Lambda_a{}^b\big)
      +\delta_G\big(\Lambda^a\big) +\delta_Z\big(\Sigma\big)
+\delta_+\big(\Upsilon_+\big)
      +\delta_-\big(\Upsilon_-\big) \,,
\end{align}
where the parameters of the transformations on the r.h.s.~are given by
  \begin{align}\begin{split}\label{softalgparam}
 \Xi^\mu &= \frac12\,\bar\epsilon_{2+}\gamma^0\epsilon_{1+}\,\tau^\mu
   +\frac12\big(\bar\epsilon_{2+}\gamma^a\epsilon_{1-}+\bar\epsilon_{2-}\gamma^a\epsilon_{1+}\big)e^\mu{}_a \,, \\[.1truecm]
 \Lambda^{ab} &= -\Xi^\mu\omega_\mu{}^{ab}
-S\,\bar\epsilon_{2+}\gamma^{ab}\epsilon_{1+} \,, \\[.1truecm]
 \Lambda^a &= -\Xi^\mu\omega_\mu{}^a
             +S\big(\bar\epsilon_{2+}\gamma^{a0}\epsilon_{1-} +\bar\epsilon_{2-}\gamma^{a0}\epsilon_{1+}\big)\,, \\[.1truecm]
 \Upsilon_\pm &= -\Xi^\mu\psi_{\mu\pm} \,, \\[.1truecm]
 \Sigma &= -\Xi^\mu m_\mu +\bar\epsilon_{2-}\gamma^0\epsilon_{1-}
\,.
\end{split}\end{align}
We derive the following chain of constraints by supersymmetry variations
\begin{align}
  \stackrel{Q_-}{\longrightarrow} \hskip1.1cm \hat\psi_{ab-} &=0 \label{offmoreconstraints}\\
 \hat R_{\mu\nu}(H)=0 \quad \stackrel{Q_+}{\longrightarrow} \quad \hat\psi_{\mu\nu+}=0 \quad
     \stackrel{Q_+}{\longrightarrow} \quad
 \hat R_{\mu\nu}{}^{ab}(J) &=-4\,\varepsilon^{ab}\,\tau_{[\mu}\hat D_{\nu]}\,S \,. \label{offconstraints}
\end{align}
Note that this is a subset of the constraints given in eqs.~\eqref{moreconstraints}--\eqref{onshell}.
The Bianchi identities, upon use of the last constraint in eq.~\eqref{offconstraints}, get the following contributions from
the auxiliary field:
\begin{align}\label{offBianchi}
  \hat R_{0[a}{}^{b]}(G) = 0 \,, \hskip2cm
  \hat R_{ab}{}^c(G) = 2\,\varepsilon_{ab}\,e^{\mu c}\,\hat D_{\mu}\,S \,.
\end{align}
With the help of these identities one can show that the supersymmetry variation of the constraint
given in eq.~\eqref{offmoreconstraints} does not imply
any further constraints. The contribution of the auxiliary field in the
last equation of \eqref{offconstraints} ensures that its variation
does not lead to additional constraints. The set of constraints given
in eqs.~\eqref{offmoreconstraints}--\eqref{offBianchi} is thus complete
because we varied \emph{all} constraints under supersymmetry.  The
check here is more complete than in the on-shell case where we did not
vary the bosonic equation of motion anymore. Since the on-shell case
can be derived from the off-shell formulation, see below, we have also
proven consistency of the on-shell formulation.

As a consistency check we note that the above result for the off-shell
multiplet entails the two on-shell formulations that were presented
earlier in the literature. First of all, by imposing
\begin{align}
 S=0 \,,
% \hskip2cm \gamma^a\hat\psi_{a0-}=0 \,,
\end{align}
we arrive at the result of section \ref{subsec:onshellNC}. Secondly,  choosing
\begin{align}
 S=\frac{1}{2R} \,,
% \hskip2cm \gamma^a\hat\psi_{a0-}=0 \,,
\end{align}
with $R$ constant, and related to the cosmological constant by $\Lambda_{CC}=-1/R^2$, we reproduce the on-shell
Newton--Hooke supergravity theory of \cite{Bergshoeff:2014gja}. The $1/R^2$ corrections w.r.t.~the flat case are
hidden in the curvatures, e.g.~the bosonic equation of motion for Newton--Hooke supergravity is still given by
eq.~\eqref{onshell}, but $\hat R_{\mu\nu}{}^a(G)$ now contains additional  terms of order $1/R$.

This concludes the discussion of the off-shell formulation of non-relativistic three-dimen\-sion\-al Newton--Cartan
supergravity.

%%%%%%%%%%%%%%%%%%%%%%%%%%%%%%%%%%%%%%%%%%%%%%%%%%%%%%%%%%%%%%%%%%%%%%%%%%%%%%%%%%%%%%%%%%%%%%%%%%%%%%%%%%%%%
%%%%%%%%%%%%%%%%%%%%%%%%%%%%%%%%%%%%%%%%%%%%%%%%%%%%%%%%%%%%%%%%%%%%%%%%%%%%%%%%%%%%%%%%%%%%%%%%%%%%%%%%%%%%%

\section{Conclusions and outlook}\label{sec:conclusions}

In this paper, we have discussed how generally covariant,
non-relativistic (super)gravity theories can be obtained from
relativistic ones via a procedure that implements the
non-relativistic limit. The method extends the In\"on\"u--Wigner
contraction, that yields a non-relativistic space-time symmetry algebra
starting from a relativistic one, to an irreducible (super)multiplet of fields representing the algebra.
In applying this method, special care has to be taken of various
consistency checks to avoid divergences in geometric quantities and
transformation rules. We have shown how this procedure can be used to
obtain torsionless Newton--Cartan gravity from General Relativity,
three-dimensional on-shell Newton--Cartan supergravity from
relativistic on-shell supergravity and how it can be used to obtain
non-relativistic superparticle actions from relativistic ones.

We would also like to remind the reader that if we are not interested in \emph{super}gravity we are free to impose the weaker
`kinematical' constraint \eqref{offMcurv}, which might lead to twistless torsionfull Newton--Cartan structures. However, also in
this case we do get restrictions on the gauge-field $\tau_\mu$, see eqs.~\eqref{offRhconstr} and \eqref{TTNC??}. In contrast, the
limit discussed in \cite{Jensen:2014wha} does not lead to any restriction of $\tau_\mu$.

In contrast to methods that are based on the gauging of algebras, the
limiting procedure has the advantage that it can be extended to the
case in which the relativistic (super)multiplet contains (auxiliary)
fields that are not associated to gauge-fields of the underlying
space-time symmetry algebra. As an example, we have derived a new
off-shell formulation of three-dimensional Newton--Cartan
supergravity containing a real auxiliary scalar $S$.

Several extensions of this work can be considered.
For example, now that  the general limiting procedure has been defined, it would be interesting to apply it to
a specific version of off-shell 4D ${\cal N}=2$ Poincar\'e supergravity to obtain
4D off-shell Newton--Cartan supergravity.
We note that the examples considered in this paper only dealt with pure Newton--Cartan
(super)gravity theories. It would be interesting to extend the
procedure to matter coupled (super)gravity theories (see
\cite{Jensen:2014wha,Geracie:2015xfa} for similar ideas applied to
condensed matter theories). Another interesting extension is to
consider relativistic theories whose underlying symmetry algebra is
different from the Poincar\'e (super)algebra, such as conformal
ones. In particular, one can define a contraction from the
relativistic (super)conformal algebra to the Galilean (super)conformal
algebra, that has been discussed in the context of non-relativistic
limits of AdS/CFT \cite{Bagchi:2009my} and flat space holography (see
e.g. \cite{Bagchi:2010zz}).  One could try to extend this contraction
to a vielbein (super)multiplet of the (super)conformal algebra and in
this way find background theories for (supersymmetric versions of) the
Galilean conformal algebra, see
e.g.~\cite{Beckers:1986ty,Beckers:1987xr,Gauntlett:1990xq}. Note that
it is not obvious how to do this via a gauging procedure, as the
Galilean conformal algebra does not allow for the type of central
extension that was crucial in the gauging of the Bargmann algebra.

As we mentioned in the introduction this work only considers 'particle' backgrounds. It would be
interesting to extend the
limiting procedure to the case of non-relativistic branes. Here, one would first have to find a
suitable extension of the Poincar\'e algebra whose contraction leads to extended stringy Galilei
algebras \cite{Brugues:2004an}.

Another interesting extension is to consider
 other limits than the non-relativistic one, such as
the ultra-relativistic limit.\footnote{We thank Quim Gomis to suggest this extension to us.} At the level of algebras, the latter
yields a contraction of the Poincar\'e algebra to the Carroll algebra. It would
be interesting to see whether  this algebra contraction can be extended to a  relativistic vielbein
multiplet and to check whether such a limit can e.g.~be used to derive the
recently constructed action for a Carroll (super-)particle in a
curved background \cite{Bergshoeff:2014jla,Bergshoeff:2015wma}.

We should stress that, as presented in this work, it is not clear whether  every algebra contraction can be translated into a
contraction at the level of the field theory representing that algebra. Moreover, certain non-relativistic symmetry algebras
cannot be viewed as contractions of relativistic ones. An example of such an algebra is given by the Schr\"odinger algebra.
On the other hand, the Bargmann and the Schr\"odinger algebra can be obtained as light-like reductions of relativistic algebras
\cite{Duval:1984cj,Duval:1990hj}. Perhaps one can define a different sort of contraction or limiting procedure related to such kind of
reductions which would give rise to (torsional) Newton--Cartan structures as presented in \cite{Andringa:2010it,Bergshoeff:2014uea}. 
In view of the recent applications of torsional Newton--Cartan geometry in non-relativistic holography
\cite{Christensen:2013lma,Christensen:2013rfa,Hartong:2014oma,Hartong:2014pma,Hartong:2015wxa},
it would be interesting to investigate this case in more detail.

\section*{Acknowledgements}

It is a pleasure to thank Hamid Afshar, G\"okhan Alka\c c, Joaquim Gomis, Jelle Hartong, Marija Kova\v cevi\'c,
Lorena Parra and Blaise Rollier for discussions.
JR acknowledges financial support by the START project Y 435-N16 of the Austrian Science Fund (FWF)
and TZ was supported by the Dutch Academy of Sciences (KNAW).

%%%%%%%%%%%%%%%%%%%%%%%%%%%%%%%%%%%%%%%%%%%%%%%%%%%%%%%%%%%%%%%%%%%%%%%%%%%%%%%%%%%%%%%%%%%%%%%%%%%%%%%%%%%%%%%%
%%%%%%%%%%%%%%%%%%%%%%%%%%%%%%%%%%%%%%%%%%%%%%%%%%%%%%%%%%%%%%%%%%%%%%%%%%%%%%%%%%%%%%%%%%%%%%%%%%%%%%%%%%%%%%%%

% \appendix
%
%
%

%%%%%%%%%%%%%%%%%%%%%%
%%%%%%%%%%%%%%%%%%%%%%              Cut here
%%%%%%%%%%%%%%%%%%%%%%

% \bibliographystyle{utphys}
% \bibliography{Newton-Cartan}
\small{

\providecommand{\href}[2]{#2}\begingroup\raggedright\endgroup

}

\end{document}